\documentclass[twocolumn]{aastex702}
\usepackage[T1]{fontenc}
\usepackage{booktabs,amsmath,array}
\usepackage{xcolor}
\usepackage{placeins}
\hypersetup{hypertexnames=false}
\newcommand{\alttext}[1]{}
\shorttitle{Agentic Observer Hackathon}
\shortauthors{Wang et al.}
\begin{document}
\title{The Agentic Observer Hackathon:\\ Autonomous Decision-making in Simulated Astronomical Surveys}
\correspondingauthor{Ting Tan and Huanyuan Shan}
\author{Haosong Wang}
\affiliation{Shanghai Astronomical Observatory, Chinese Academy of Sciences, Shanghai 200030, P. R. China}
\affiliation{University of Chinese Academy of Sciences, Beijing 100049, P. R. China}
\email{}

\author{Ting Tan}
\affiliation{IRFU, CEA, Universit\'e Paris-Saclay, F-91191 Gif-sur-Yvette, France}
\affiliation{Shanghai Astronomical Observatory, Chinese Academy of Sciences, Shanghai 200030, P. R. China}
\email[show]{ting.tan.astro@outlook.com}

\author{Yao Li}
\affiliation{Upstream Labs (Shenzhen Upstream Labs Technology Co., Ltd), Room 505M, 5th Floor, Building 522, Bagualing Industrial Zone, Bagua 3rd Road, Hualin Community, Yuanling Subdistrict, Futian District, Shenzhen, China}
\email{}

\author{Huanyuan Shan}
\affiliation{Shanghai Astronomical Observatory, Chinese Academy of Sciences, Shanghai 200030, P. R. China}
\email[show]{hyshan@shao.ac.cn}

\author{Dezi Liu}
\affiliation{South-Western Institute for Astronomy Research, Yunnan University, Kunming 650500, P. R. China}
\email{}

\author{Xiangkun Liu}
\affiliation{South-Western Institute for Astronomy Research, Yunnan University, Kunming 650500, P. R. China}
\email{}

\author[0000-0001-7729-6629]{Yifei Luo}
\affiliation{Lawrence Berkeley National Laboratory, 1 Cyclotron Road, Berkeley, CA 94720, USA}
\email{}

\author{Wei Wang}
\affiliation{State Key Laboratory of Dark Matter Physics, Shanghai Key Laboratory for Particle Physics and Cosmology, Key Laboratory for Particle Astrophysics and Cosmology (MOE), Physics-AI Institute, School of Physics and Astronomy, Shanghai Jiao Tong University, Shanghai 200240, China}
\affiliation{Southern Center for Nuclear-Science Theory (SCNT), Institute of Modern Physics, Chinese Academy of Sciences, Huizhou 516000, Guangdong Province, China}
\email{}

\author{Zonghuan Wu}
\affiliation{Relevant Study LLC, 10749 Brookwell Drive, Cupertino, CA 95014, USA}
\email{}

\author{Zhongxu Zhai}
\affiliation{State Key Laboratory of Dark Matter Physics, Tsung-Dao Lee Institute \& School of Physics and Astronomy, Shanghai Jiao Tong University, Shanghai 200240, China}
\affiliation{Shanghai Key Laboratory for Particle Physics and Cosmology, and Key Laboratory for Particle Physics, Astrophysics and Cosmology, Ministry of Education, Shanghai Jiao Tong University, Shanghai 200240, China}
\affiliation{Waterloo Center for Astrophysics, University of Waterloo, Waterloo, ON N2L 3G1, Canada}
\affiliation{Department of Physics and Astronomy, University of Waterloo, Waterloo, ON N2L 3G1, Canada}
\email{}

\begin{abstract}
Astronomical surveys must allocate observing time while responding to changing conditions and competing scientific priorities. We present the Agentic Observer Hackathon (AOH), a community challenge for studying autonomous survey operations. Within a synthetic multi-object survey with prescribed exposure responses and assignment constraints, decision systems select pointings, assign targets and choose exposure configurations while responding to observing requests, instrumental faults and data loss. We evaluate systems submitted by 102 teams on campaigns reserved for evaluation and analyse target completion in a reference campaign. The leading systems combine numerical exposure planning with language models for interpreting operational information and supporting longer-term decisions. Exploratory tests show improved fault reporting with model access in several systems, while revised scientific priorities produce substantial changes in ranking. The reference campaign traces target completion and diagnoses errors in the baseline's inference of prescribed quality bands. The released simulator and reference strategies provide common survey conditions for comparing exposure allocation, target selection and recovery from interruptions.
\end{abstract}

\section{Introduction}
\label{sec:intro}

Astronomical surveys allocate finite observing time among targets with different exposure requirements and scientific priorities. Each observation contributes to the survey sample and changes the targets that remain to be observed. The choice of the next field and exposure duration therefore depends on survey progress, target visibility and the prevailing observing conditions. Over a campaign, these choices determine the depth, sky coverage and completeness of the resulting sample \citep{Burleigh2020,Bianco2022,schlafly2023operations}.

Operational surveys couple scheduling to measured conditions. The Rubin scheduler was designed to balance survey objectives \citep{naghib2019scheduler,Bianco2022}, DES uses the \texttt{obstac} scheduler to select exposures from survey progress and observing conditions \citep{neilsen2019obstac}, and the Legacy Surveys adjust exposure times as conditions change \citep{Burleigh2020}. DESI selects fields, assesses exposure completeness and updates target lists automatically \citep{schlafly2023operations}. AstroSA and DeepSurveySim provide environments for comparing scheduling methods \citep{xie2024astrosa,voetberg2023deepsurveysim}. Offline reinforcement-learning studies have also compared deep Q-network scheduling variants on simulated Stone Edge Observatory campaigns \citep{terranova2023selfdriving}.

Language-model agents combine reasoning with actions \citep{Yao2023}, and benchmarks such as ScienceWorld and AgentBench evaluate their behaviour in interactive environments \citep{wang2022scienceworld,liu2024agentbench}. These capabilities make language-model agents promising for astronomical observation planning, where decisions must respond to changing conditions. StarWhisper Telescope integrates observation planning, telescope control and image processing, with astronomer review of observing plans \citep{wang2025starwhisper}. FORMA uses a multi-agent workflow to verify classifications and redshifts in astronomical spectra \citep{wang2026forma}. Alongside these applications, shared evaluation environments would help compare the capabilities of agentic systems on astronomical tasks. For survey operations, a common simulation would allow observing decisions to be evaluated under the same catalogues, weather histories and events. The use of shared simulations for method evaluation has precedents in astronomy, including GREAT3 and SKA Science Data Challenge 3a, which provide common datasets and assessment procedures \citep{mandelbaum2014great3,bonaldi2025sdc3a}.

\begin{figure*}[ht!]
\centering
\includegraphics[width=\textwidth]{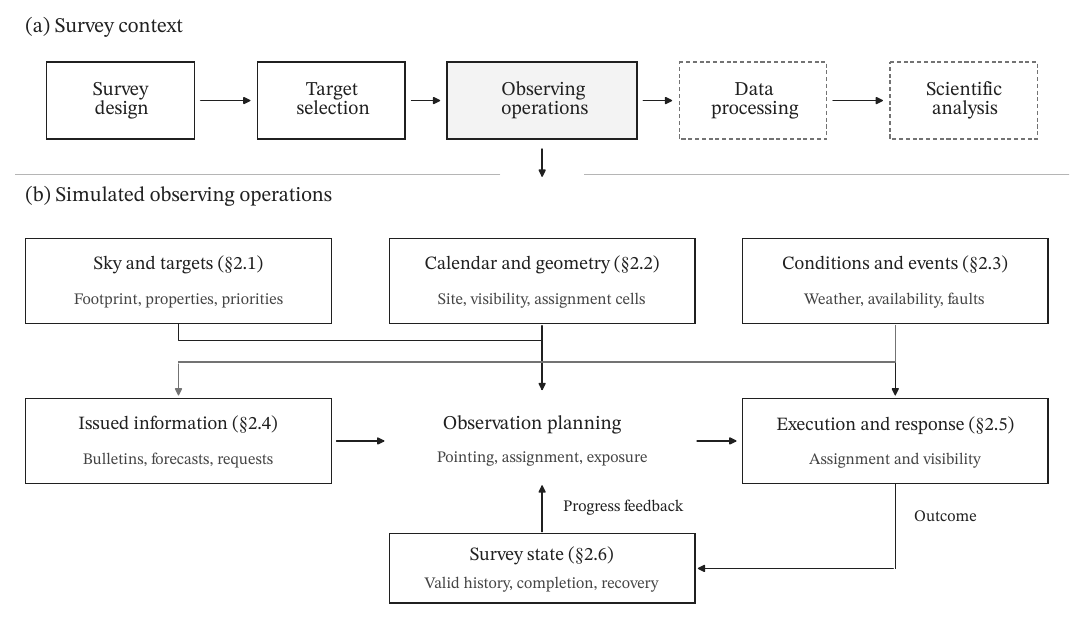}
\caption{Survey operations represented in AOH. (a) Stages of a survey, with the simulated observing operations highlighted and the preceding stages supplying fixed task inputs. (b) The six simulation modules, labelled by subsection, and their relationships to observation planning. Arrows show inputs, execution and feedback. Dashed boxes identify downstream survey stages outside the simulated scope; solid boxes in panel (b) denote implemented components. The two-level layout is visually inspired by Figure~5 of \citet{mandelbaum2014great3}.}
\label{fig:operations}\label{fig:overview}
\end{figure*}

We present the Agentic Observer Hackathon (AOH), a community challenge in which participants develop decision systems for a synthetic simulation of multi-object spectroscopic observations. The simulation combines field selection, target assignment and exposure planning with changing weather, observing requests and instrumental events (Figure~\ref{fig:operations}). We evaluate 102 submitted systems on four held-out campaigns and examine their survey outcomes and run-to-run variability. The leading systems combine numerical exposure planning with language models for interpreting operational information and supporting longer-term decisions. A reference campaign relates exposure choices and instrument performance to target completeness, while exploratory tests examine fault reporting and adaptation to revised scientific priorities. We release the simulator, reference implementations and public campaign data to support future survey planning and the development of observing decision systems.

Section~\ref{sec:environment} describes the simulated survey and its response models. Section~\ref{sec:benchmark} presents the task design, decision interface, evaluation objective and reference methods. Section~\ref{sec:challenge} reports survey outcomes, comparisons across held-out campaigns, system designs and exploratory tests. Sections~\ref{sec:discussion} and~\ref{sec:conclusions} discuss the implications for survey operations and summarise the conclusions.

\section{Simulation of survey operations}
\label{sec:environment}
For AOH, we define a common survey environment with shared rules for observing geometry, target assignment and exposure response. All systems evaluated on a given task receive the same catalogue and observing calendar, and encounter the same weather and event histories. We model coupled field selection, target assignment and exposure planning in a synthetic multi-object spectroscopic survey. DESI provides context for the target classes and the organisation of survey operations \citep{desi2016science,schlafly2023operations}. Each task specifies a sky footprint, target catalogue, observing calendar and assignment grid. Atmospheric conditions, instrumental performance and operational events evolve during the campaign (Figure~\ref{fig:operations}). The target properties and observing conditions determine each exposure response, and the valid observation history determines survey progress.

Table~\ref{tab:model_basis} summarises the models and evaluation choices used to compare observing strategies. Target strengths and observing conditions set exposure requirements (Sections~\ref{sec:targets} and~\ref{sec:response}), while science weights, penalties and operational rewards define the campaign objective (Section~\ref{sec:objective}). Decision systems use bulletins, forecasts and exposure feedback to revise their plans as conditions change, new requests arrive or operational events affect observations (Sections~\ref{sec:conditions} and~\ref{sec:information}). Note that the response, weather and scoring coefficients are set specifically for the challenge. Adapting the simulation to real surveys requires calibrating target responses against spectra and exposure measurements, and weather and instrument models against site monitoring and operational records.

\begin{table*}[tp]
\centering\small\setlength{\tabcolsep}{5pt}
\caption{Astronomical basis and prescribed choices in the AOH simulation.}
\label{tab:model_basis}
\begin{tabular}{>{\raggedright\arraybackslash}p{3.0cm}>{\raggedright\arraybackslash}p{3.1cm}>{\raggedright\arraybackslash}p{9.3cm}}
\toprule
Component & Basis & Role in the simulation\\
\midrule
Sky geometry & Astronomical methods & Computes visibility and airmass within prescribed observing limits and calendars.\\
Target classes and feature strength & Prescribed models & Class fractions set the DESI-inspired target mix; $f_i$ sets relative exposure requirements.\\
Exposure quality and response & Prescribed models & Maps target strength, exposure duration and conditions to the capped response $g_{ie}$.\\
Weather and disturbances & Prescribed models & Generates seasonal and correlated weather, closures and changes in instrument efficiency.\\
Class weights and completion & Evaluation choices & Weights $w_i$, required status and penalties set priorities; a response threshold defines completion.\\
Programme matching & Evaluation choices & Band thresholds and multipliers reward selection of prescribed quality labels.\\
Fault and request rewards & Evaluation choices & Rewards correct fault reports and fulfilled requests; penalises false reports.\\
\bottomrule
\end{tabular}
\tablecomments{Astronomical methods use established geometry and analytic approximations. Prescribed models specify synthetic target populations, exposure responses and observing conditions. Evaluation choices define priorities, completion criteria and rewards.}
\end{table*}

\begin{figure*}[ht!]
\centering
\includegraphics[width=\textwidth]{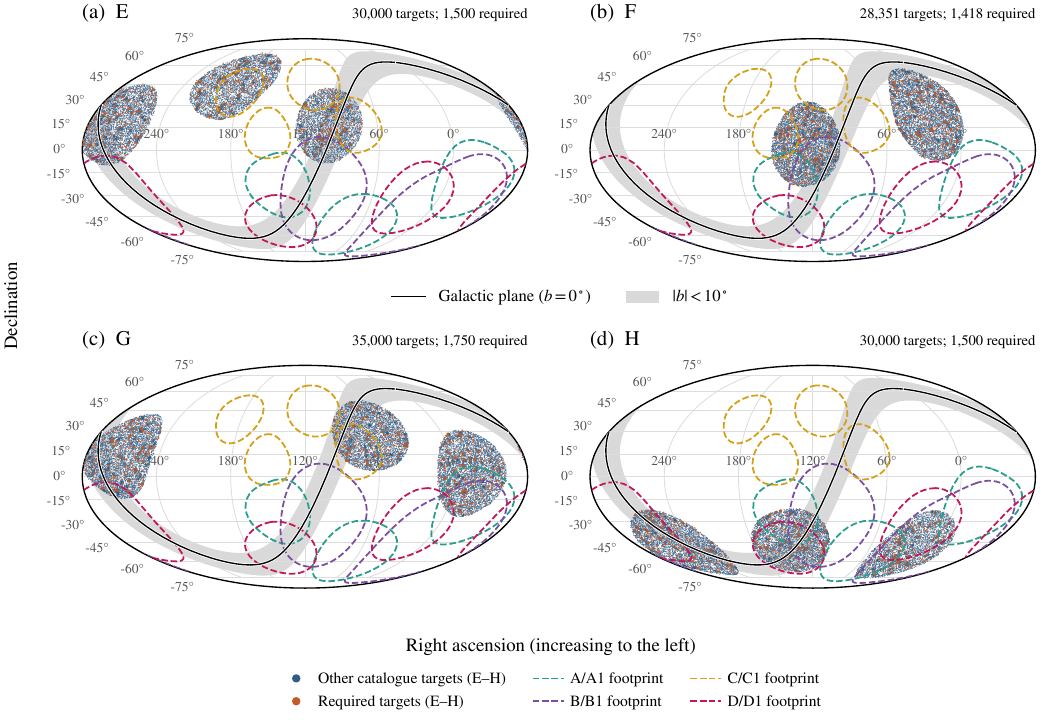}
\caption{Sky distributions of the final test catalogues E--H in an equal-area Mollweide projection. Orange points show the randomly selected required subset; blue points show the remaining catalogue targets. All targets contribute to the science score. The score also penalises required targets that remain incomplete at termination. Each panel gives the total and required-target counts. The coloured dashed outlines show the public A--D footprints, shared with A1--D1. The black solid curve marks the Galactic plane ($b=0^\circ$). Grey shading marks $|b|<10^\circ$ as a reference for proximity to the Galactic plane. Right ascension increases to the left, with a central meridian of $120^\circ$. The E--H catalogues come from the registered final scenario bundles; the public footprints were verified against the release on 2026 October 10.}
\label{fig:catalogues}
\alttext{Four sky maps show the final catalogues E, F, G and H. Orange points mark required targets, blue points mark catalogue targets outside the required subset, and four coloured dashed outlines show the public A--D footprints. A black solid curve in each panel marks the Galactic plane. Light grey shading shows the illustrative region within 10 degrees of the Galactic plane.}
\end{figure*}

\subsection{Sky coverage and target population}
\label{sec:targets}
We include several target populations to compare how observing time is shared among objects with different exposure requirements and scientific priorities. The catalogue adopts the broad classes used in DESI target selection \citep{myers2023targeting,hahn2023bgs,zhou2023lrg,raichoor2023elg,chaussidon2023qso}. It contains stars and four extragalactic populations: bright, predominantly low-redshift galaxies from the Bright Galaxy Survey (BGS); luminous red galaxies (LRG), dominated by older stellar populations; star-forming emission-line galaxies (ELG); and quasars (QSO), luminous accreting galactic nuclei.

For an observed target, the exposure needed to reach a specified signal-to-noise ratio depends on its apparent magnitude and spectrum, the telescope collecting area, instrumental throughput, background and observing conditions \citep{schlafly2023operations}. In the simulation, the dimensionless feature strength $f_i$ sets the exposure $T_0f_0/f_i$ needed to reach unit response at mean observing quality $q=1$, where $f_0$ and $T_0$ define the exposure scale (Section~\ref{sec:response}). Larger strengths therefore require shorter exposures. We draw $f_i$ from class-dependent lognormal distributions, which give positive strengths and a continuous range of exposure requirements. Their prescribed medians give stars and BGS galaxies shorter reference exposures than LRGs, ELGs and quasars (Table~\ref{tab:populations_efgh}).

Each target also has a prescribed science weight $w_i$, which sets its contribution to the survey score. Higher weights for quasars and lower weights for stars and BGS galaxies create a trade-off between exposure demand and adopted scientific priority (Table~\ref{tab:populations_efgh}). Class priorities also enter DESI target assignment \citep{schlafly2023operations}. Our strength distributions and weights are prescribed for the challenge; a real survey requires strengths calibrated against its spectra and exposure measurements, and weights set by its scientific objectives. A required subset is selected independently of class, strength and sky position. A penalty for required targets that remain incomplete gives this subset an additional completion priority.

We construct the survey footprints from separated spherical polygons with irregular boundaries and specified areas. Their locations and areas vary between tasks, giving different seasonal visibility from the observing sites (Section~\ref{sec:tasks}). Each catalogue contains a diffuse population and angular overdensities. These density variations and irregular boundaries make field placement affect how many targets can be assigned in an exposure. We retain targets with an observing window during the survey season. Figure~\ref{fig:catalogues} shows the final test catalogues E--H and the public footprints; Appendix~\ref{app:generators} describes their construction.

\subsection{Observing windows and target assignment}
The observing site and season set the accessible sky and the available night time. We use astronomical darkness, with solar altitude below $-18^\circ$, and require targets to remain above $30^\circ$ throughout an exposure. The reference catalogue selection requires at least one continuous 900-s window satisfying these limits during the season. Thus every accepted target has a geometric observing opportunity, while weather and competition for observing time determine whether it is completed. Requested exposures last 60--3600\,s; those starting within a night window end at the earliest of the requested end, the end of that night and the end of the campaign. Exposure and waiting share the available night time; slews, target reassignment and readout have zero simulated duration. The calendar construction and treatment of gaps are specified in Appendix~\ref{app:conventions}.

We use a square grid of $n^2$ cells on the tangent plane to impose finite spatial capacity: one target per cell per exposure. Nearby targets compete for cells, so dense regions can require several visits. The final E--H tasks use 9--100 cells (Table~\ref{tab:operations}). For comparison, DESI uses 5000 robotic fibre positioners with finite, overlapping patrol regions and collision constraints \citep{desi2022instrument,silber2023focalplane}. These constraints affect assignment completeness \citep{smith2019fiberassignment}. In AOH, the commanded altitude--azimuth pointing defines the initial field centre, including any scenario pointing offset, and the field then follows sidereal tracking. At exposure start, the grid axes follow increasing local altitude and azimuth. Target-to-cell assignments and the grid orientation relative to the tracked field are held fixed during that exposure. Targets must remain above the altitude limit throughout. Appendix~\ref{app:conventions} gives the projection and geometric eligibility conditions.

\begin{figure*}[ht!]
\centering
\includegraphics[width=0.94\textwidth]{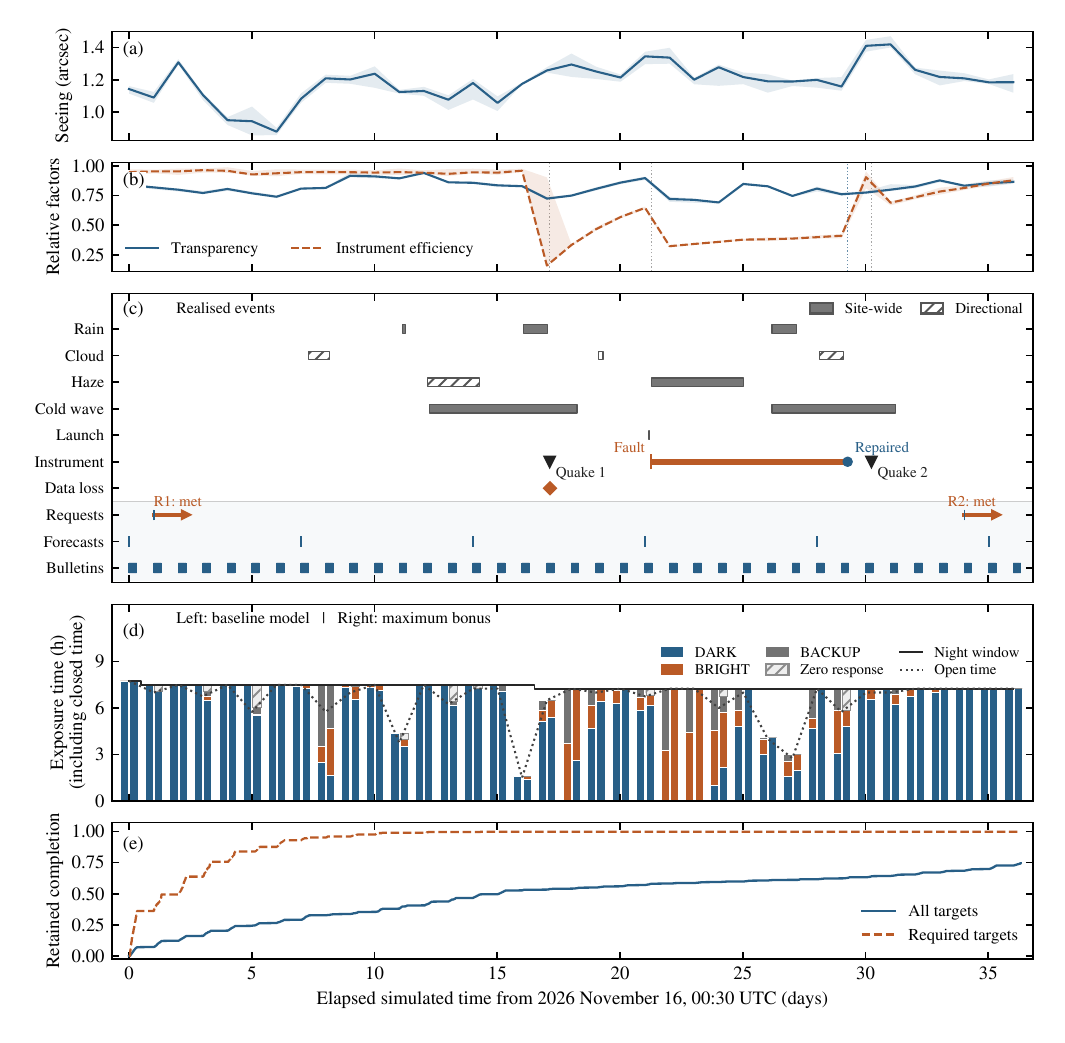}
\caption{Evolution of the 37-night L3 campaign with the baseline model (Section~\ref{sec:baselines}) and language-model advice disabled. (a,b) Nightly median seeing, transparency and instrument efficiency over open intervals; shading shows the interquartile ranges. Vertical guides mark the instrumental events and fault repair. (c) Realised events above the divider and released information below it. Grey bars mark weather and launch events, with hatching for directional effects. Triangles mark earthquakes, the orange bar the active fault, the blue circle its repair, and the diamond the data-loss notification. Blue ticks mark information issue times; orange windows mark observing requests. (d) Nightly exposure time by programme. Left bars show the programmes selected by the baseline model; right bars show the programme giving the largest weighted bonus for each exposure. The reference uses the full simulated conditions at fixed pointing, assignments and duration (Appendix~\ref{app:reference_reconstruction}). Hatched segments denote zero-response ties. The solid and dotted lines show night-window and open time, respectively. Exposure-action time includes closed intervals and subsequently invalidated exposures. (e) Retained completion for all 8700 targets and the 435 required targets, with threshold $G_i\geq0.5$.}
\label{fig:timeline}
\alttext{Five aligned panels show nightly conditions; realised events and information releases; paired programme-time bars for the baseline model and the fixed-exposure maximum-bonus reference; and retained completion. An active-fault interval ends at its repair marker. Two earthquake markers and a data-loss marker identify other disturbances. Both request windows are marked as met. Instrument efficiency recovers after repair, and subsequent programme choices mostly agree with the reference.}
\end{figure*}

\subsection{Atmospheric conditions and operational events}
\label{sec:conditions}
We describe observing conditions by seeing, transparency, sky quality and instrument efficiency, together with intervals when the site is closed. Seeing is measured in arcseconds; the other factors are dimensionless. A larger sky-quality factor represents a more favourable background. We generate synthetic atmospheric histories with a seasonal mean and autoregressive fluctuations \citep{box2015timeseries}. Seasonal variation changes observing opportunities across the campaign, while correlated fluctuations make recent conditions informative for the next exposure. Their amplitudes and persistence set how rapidly the required exposure time and preferred programme change. All policies evaluated on a given task use the same weather and event histories. Appendix~\ref{app:weather} gives the weather and event models.

Operational events include rain, clouds, haze and cold waves, launch-related closures, terrain obstructions, earthquakes and instrumental faults. An event can affect the whole site or a horizon sector, changing target visibility or exposure quality. Earthquakes reduce instrument efficiency, which then recovers over successive nights. A correct report of an active reportable fault repairs the instrument. Tasks with additional disturbances also include pointing offsets and loss of earlier observations. These events test responses to temporary losses, report-triggered repair and replacement observations, together with adjustments to pointing. We use the local task L3 as a diagnostic example because its complete scenario products and archived actions allow us to follow conditions, decisions and target completion through observing requests, instrument losses and data loss. Figure~\ref{fig:timeline} shows this sequence; Section~\ref{sec:localdiagnostic} analyses the survey outcome, and Appendix~\ref{app:reference_reconstruction} gives the event timings and effects.

\subsection{Observing information and time-dependent requests}
\label{sec:information}
The decision system receives bulletins, forecasts and exposure feedback as the campaign advances. Bulletins identify event categories and approximate directions, forecasts describe upcoming weather and launch events, and exposure outcomes provide information about instrument performance. Records become available at their assigned issue times; observing decisions therefore use the catalogue, information released so far and previous outcomes. Tasks A1--D1 and E--H additionally provide free-text duty logs. 

Time-sensitive observing programmes allocate telescope time in response to external triggers \citep{andreoni2024rubintoo}. We represent changing priorities through observing requests with deadlines, making the value of an exposure depend on when it is taken. Each request specifies its targets, issue time, deadline, response threshold, minimum completed-target count and reward. To satisfy a request, the relevant valid exposures must lie wholly within its issue-to-deadline window. These requests compete with the ongoing survey for the remaining observing time, and their completion can change if supporting observations are later lost. Section~\ref{sec:benchmark} gives the scoring rules; Appendix~\ref{app:weather} describes request generation and the reference information cadence.

\subsection{Exposure response and programme matching}
\label{sec:response}
For faint, background-limited point-source imaging, effective exposure scales with atmospheric transmission squared and inverse PSF FWHM squared at fixed sky brightness \citep{neilsen2016quality}. DESI computes effective time using reference spectra and source profiles, with seeing entering through the fraction of light captured by a fibre \citep{schlafly2023operations,guy2023pipeline}. AOH uses a prescribed response with first-power transparency and inverse first-power seeing. The challenge response grows linearly with condition-weighted exposure up to unity.

For target $i$ in exposure $e$, we define the prescribed mean observing quality as
\begin{equation}
q_{ie}=\frac{1}{q_0 T_e}
\sum_{p\in\mathcal P_e}\Delta t_p D_{ip}
\frac{\eta_{ip}\tau_{ip}\kappa_{ip}L_i(t_p)}
{s_{ip}X_i(t_p)^\beta},
\label{eq:quality}
\end{equation}
where $T_e$ is the actual exposure duration and $q_0$ is a dimensionless normalisation. The factors $\eta$, $\tau$ and $\kappa$ describe instrument efficiency, transparency and sky quality; $s$ is seeing in units of $1\,\mathrm{arcsec}$, $X$ is airmass, $\beta$ is the airmass exponent, and $L$ is the lunar factor. The indicator $D$ is unity for an available sightline and zero during closure or obstruction. We divide the exposure into intervals $\mathcal P_e$ of duration $\Delta t_p$ and evaluate the factors at their midpoints $t_p$, with $\sum_p\Delta t_p=T_e$. All factors include the relevant events and repairs. Closed or obstructed intervals have zero response and contribute to the elapsed duration $T_e$. The product $T_eq_{ie}$ is the challenge effective exposure used below. Within each open interval, efficiency, transparency and sky quality multiply the time contribution, while seeing and the airmass factor enter in the denominator.

The airmass uses the functional form of \citet{kasten1989airmass}, evaluated with the simulator's geometric zenith angle and normalised at the zenith (Appendix~\ref{app:response_integration}). Physical scattered-moonlight models calculate a contribution that adds to the sky background \citep{krisciunas1991moonlight,jones2013moonlight}. In AOH, $L$ is a dimensionless multiplicative quality penalty determined by lunar illumination, altitude and target--Moon separation. Its amplitude and angular scale are challenge parameters. The normalisation $q_0$ sets the exposure scale, while $\beta$ controls the airmass dependence. Appendix~\ref{app:conventions} specifies these functions and the numerical integration, and Table~\ref{tab:simulation} lists the shared settings.

For a geometrically eligible assignment, the raw response is
\begin{equation}
g_{ie}=\min\left(\frac{f_iT_eq_{ie}}{f_0T_0},1\right),
\label{eq:response}
\end{equation}
where $f_0$ and $T_0$ are the reference strength and exposure time for the task. A target with $f_i=f_0$ reaches $g_{ie}=1$ after an effective exposure $T_0$. Capping the response gives each target a finite return, creating a trade-off between deepening an exposure and moving to other targets. For a desired response $0<g_\star\leq1$, equation~\ref{eq:response} gives the condition $T_eq_{ie}\geq g_\star f_0T_0/f_i$. At fixed $q_{ie}>0$, this corresponds to a duration threshold $T_\star=g_\star f_0T_0/(f_iq_{ie})$. The requested duration is an allowed integer number of seconds, and the actual exposure must reach $T_\star$ within the observing window. When conditions change during an exposure, the effective exposure is calculated from equation~\ref{eq:quality}.

The science-weighted contribution is $c_{ie}=w_i g_{ie}m_{ie}$. DARK, BRIGHT and BACKUP are interface labels for the high, intermediate and low bands of a synthetic quality statistic. Selecting the matching label earns a fixed score multiplier $m_{ie}$ (Table~\ref{tab:simulation}). We assign the band using site atmosphere, sky quality, airmass and the lunar factor (equation~\ref{eq:band}). This makes band identification and instrument-efficiency estimation distinct parts of the challenge. Airmass and lunar separation vary across a field, so each target has its own band. Instrument efficiency and directional event effects enter the exposure response in equation~\ref{eq:quality}; at fixed site conditions and geometry, an efficiency loss can reduce the response while preserving the band. In DESI operations, observing programmes specify target populations and effective-depth requirements \citep{schlafly2023operations}.

\subsection{Survey progress and recovery}
\label{sec:ledger}
We retain the best valid responses so that each progress measure remains traceable to an exposure, including after data loss. For the set $\mathcal V_i$ of its valid exposures, the retained science contribution and raw response are
\begin{equation}
C_i=\max_{e\in\mathcal V_i}c_{ie},\qquad
G_i=\max_{e\in\mathcal V_i}g_{ie},
\label{eq:ledger}
\end{equation}
with $C_i=G_i=0$ for an empty history. The two maxima are calculated independently and can correspond to different exposures. We adopt $G_i\geq0.5$ as the challenge completion criterion: at least one valid exposure must reach half the reference response. Targets completed below saturation can gain further response up to unity. Further visits can increase $C_i$ or $G_i$ and retain a usable measurement if an earlier exposure is lost. DESI combines spectra across exposures \citep{guy2023pipeline}; Section~\ref{sec:discussion} discusses an extension based on cumulative depth.

A data-loss event invalidates a contiguous sequence of completed exposures. We recalculate $C_i$, $G_i$ and request completion from the surviving observations and return the updated progress to the decision system. A request reward is withdrawn if the remaining exposures fail its completion criterion. The time spent on lost exposures remains part of the campaign duration, and replacement observations use the remaining observing time.

\section{Task design, evaluation objective and reference methods}
\label{sec:benchmark}
We compare policies within the shared environment of Section~\ref{sec:environment} using a common decision interface and survey objective. Each task fixes the inputs and conditions for every evaluated system; tasks vary the available sky, observing season, assignment geometry and operational conditions. The objective combines retained target response with completeness, coverage and operational outcomes. Released reference methods provide common starting points and comparisons. Section~\ref{sec:evaluation} specifies the execution budgets, repeated evaluations and score aggregation.

\subsection{Task design and evaluation roles}
\label{sec:tasks}
The suite comprises five families of four tasks, separating local development, platform practice, public competition, duty-log extensions and held-out evaluation (Table~\ref{tab:task_design}). Development tasks provide complete scenario products for offline testing. Practice and competition tasks expose their static inputs and release observing information during a run. The held-out E--H scenarios remained on the evaluation side during the competition, with their catalogues and operational information supplied to frozen submissions through the prescribed interface. This separation tests policies on campaigns beyond those used for development.

\begin{table*}[ht!]
\centering\footnotesize\setlength{\tabcolsep}{5pt}
\caption{Observing campaigns, target catalogues and evaluation roles of the 20 tasks.}
\label{tab:task_design}\label{tab:families}\label{tab:cards}\label{tab:heldout}
\begin{tabular}{llrlrrr}
\toprule
Card & Site & Lat. ($^\circ$) & First night & Nights & Targets & Required\\
\midrule
\multicolumn{7}{l}{\textit{L1--L4: local development}}\\
L1 & Paranal & $-24.62$ & 2026-11-01 & 32 & 7\,800 & 390\\
L2 & Paranal & $-24.62$ & 2026-11-08 & 36 & 8\,600 & 430\\
L3 & Paranal & $-24.62$ & 2026-11-15 & 37 & 8\,700 & 435\\
L4 & Paranal & $-24.62$ & 2026-11-22 & 38 & 8\,800 & 440\\
\midrule
\multicolumn{7}{l}{\textit{$\alpha$--$\delta$: platform practice}}\\
$\alpha$ & Paranal & $-24.62$ & 2026-10-04 & 38 & 10\,000 & 500\\
$\beta$ & Paranal & $-24.62$ & 2026-10-18 & 38 & 9\,600 & 480\\
$\gamma$ & Paranal & $-24.62$ & 2026-10-09 & 38 & 9\,900 & 495\\
$\delta$ & Paranal & $-24.62$ & 2026-10-24 & 38 & 9\,400 & 564\\
\midrule
\multicolumn{7}{l}{\textit{A--D: public competition}}\\
A & Paranal & $-24.62$ & 2026-10-01 & 123 & 30\,000 & 1\,500\\
B & Cape Town & $-33.92$ & 2027-06-01 & 183 & 50\,000 & 2\,500\\
C & Mauna Kea & $+19.82$ & 2026-12-01 & 121 & 30\,000 & 1\,500\\
D & Nemo$^{b}$ & $-45.00$ & 2027-01-01 & 365 & 30\,000 & 1\,500\\
\midrule
\multicolumn{7}{l}{\textit{A1--D1: duty-log extension}}\\
A1 & Paranal & $-24.62$ & 2026-10-01 & 123 & 30\,000 & 1\,500\\
B1 & Cape Town & $-33.92$ & 2027-06-01 & 183 & 50\,000 & 2\,500\\
C1 & Mauna Kea & $+19.82$ & 2026-12-01 & 121 & 30\,000 & 1\,500\\
D1 & Nemo$^{b}$ & $-45.00$ & 2027-01-01 & 365 & 30\,000 & 1\,500\\
\midrule
\multicolumn{7}{l}{\textit{E--H: held-out evaluation}}\\
E & Ali & $+32.54$ & 2027-01-01 & 181 & 30\,000 & 1\,500\\
F & Gavdos$^{a}$ & $+34.84$ & 2027-10-01 & 183 & 28\,351 & 1\,418\\
G & Sahara$^{a}$ & $+18.94$ & 2027-06-01 & 214 & 35\,000 & 1\,750\\
H & Kunlun$^{a}$ & $-80.42$ & 2027-04-20 & 128 & 30\,000 & 1\,500\\
\bottomrule
\end{tabular}
\tablecomments{L1--L4 provide complete scenario products for local development. Practice and competition tasks provide static inputs and sequential observing information; A1--D1 add duty logs. E--H supply new campaigns to frozen submissions. First night gives the calendar's night-date label; observing intervals are recorded in UTC. L3 begins at 2026 November 16, 00:30 UTC. Values are from the archived task products, using the A1--D1 versions active at the close of the competition. Assignment and operational settings are listed in Table~\ref{tab:operations}. $^{a}$Simulated site at the stated coordinates. $^{b}$Fictional site.}
\end{table*}

Across tasks, changes in site and season alter target visibility and observing time, while catalogue size and assignment geometry change the demand on each field and exposure. The public A--D tasks span two to four footprint components and approximately 4000--8000\,deg$^2$. A1--D1 share the corresponding catalogues, footprints and night calendars, while adding duty-log records and using different realised weather, events and observing requests. Their comparison therefore includes changes in information, observing conditions and time-dependent priorities.

The final E--H tasks contain 28\,351--35\,000 targets over 128--214 nights, with approximately 5\% designated as required. They combine different sites, seasons and assignment geometries (Figure~\ref{fig:catalogues}). Task G pairs a small assignment field with the exposure normalisation of public task B; task H is a winter campaign at high southern latitude with pointing offsets and data loss enabled. All four include observing requests, instrumental faults and short duty-log notes. These configurations exercise the allocation of time and spatial capacity, the use of operational information and recovery under new campaign conditions. Table~\ref{tab:task_design} gives the campaign and catalogue parameters; Appendix~\ref{app:task_details} gives the assignment and operational settings.

\subsection{Decision interface and available information}
\label{sec:interface}
At each observing decision, the decision system allocates the remaining observing time using the target catalogue, forecasts and feedback from earlier exposures. Initialisation supplies the public task inputs and scoring parameters; subsequent decisions use the current simulation time, released bulletins and forecasts, new messages, request progress and previous action outcomes. The system can choose a pointing, target-to-cell assignments, observing programme and exposure duration, wait for a later opportunity, report an instrumental fault, or end the campaign. Observations and waiting advance simulated time, with zero transition overhead between exposures. Reports consume computation time and have zero simulated duration.

Exposure feedback identifies geometrically accepted targets and gives their rounded programme-weighted scores. The raw response, quality and programme multiplier are evaluated internally, so policies must infer the relevant conditions and target progress from the available feedback. The reported cumulative target utility is $\sum_i C_i$; the final score additionally accounts for completeness and operational outcomes. Every accepted action and its result are archived for reconstruction of the observing history. This defines a sequential decision problem with partial information, in which each observation both contributes to the survey and informs later choices. Appendix~\ref{app:decision} gives the formal decision-process description and feedback fields.

\subsection{Survey objective and observing trade-offs}
\label{sec:objective}
To compare observing strategies, we score each campaign by its retained target response, completion of the required subset, coverage across right-ascension bands and operational outcomes. This score defines a common set of survey priorities for the benchmark. Using $C_i$ and $G_i$ from Section~\ref{sec:ledger}, the settled campaign score is
\begin{equation}
S=\sum_i C_i-50N_{\mathrm{miss}}-200(1-J)+R_{\mathrm{req}}+R_{\mathrm{rep}},
\label{eq:total}
\end{equation}
where $N_{\mathrm{miss}}$ counts required targets with $G_i<0.5$, $J$ measures the uniformity of completion fractions across populated 10-degree right-ascension bands, $R_{\mathrm{req}}$ is the total observing-request reward, and $R_{\mathrm{rep}}$ is the net fault-report settlement. We use Jain's index for $J$ \citep{jain1984fairness}; its definition and boundary conventions are given in Appendix~\ref{app:settlement}. Using completion fractions accounts for differing numbers of catalogue targets between bands. Jain's index is unchanged when all completion fractions are multiplied by the same positive factor, so it measures their balance; the target-response term rewards the observations themselves. Two-dimensional completion maps show the finer spatial structure of the resulting sample.

The coefficients set the relative value of required completion and spatial balance. Each incomplete required target costs 50 points, while the uniformity penalty is bounded by 200. This gives required targets a strong completion incentive alongside the science-weighted return from the full catalogue. We use this scoring configuration for the reference campaign and held-out evaluation. The exploratory tasks modify selected scoring parameters (Section~\ref{sec:exploratory}). Appendix~\ref{app:settlement} compares the scales of the individual terms.

Time-dependent requests add priorities within specified intervals. A request earns its stated reward when enough requested targets reach the required response using valid exposures wholly within the issue-to-deadline window; an incomplete request contributes zero. The released cards use a response threshold of 0.5. Under the standard scoring configuration, a correct report of an active, reportable instrumental fault earns 100 and triggers repair. Two false reports are free at the start of a run and after each correct report; subsequent false reports cost 150 each until the next correct report. This allowance makes some diagnostic probes free of score penalties. The reward and penalty values set a common trade-off between detecting faults and making false reports. Data loss changes the valid target history and can withdraw completed-request rewards, as described in Section~\ref{sec:ledger}. The final objective therefore reflects both the observing plan and the effectiveness of monitoring and recovery.

\subsection{Reference methods and comparison protocol}
\label{sec:baselines}
We compare participant systems with two baseline models, denoted Baseline~1 and Baseline~2, that differ in how they estimate observing conditions and plan exposures. Each is implemented in Python, TypeScript and Rust. Baseline~1 selects exposures with a rule-based planner using target priority, remaining visibility and feedback from earlier observations. It searches candidate pointings around selected targets, fills assignment cells with accessible neighbours, selects duration using estimated gain per unit time, and chooses the observing programme from predicted target bands. Notices guide waiting and directional avoidance, and persistent response degradation can trigger a fault report. An optional language model provides advice to this planner.

\begin{figure*}[ht!]
\centering
\includegraphics[width=0.985\textwidth]{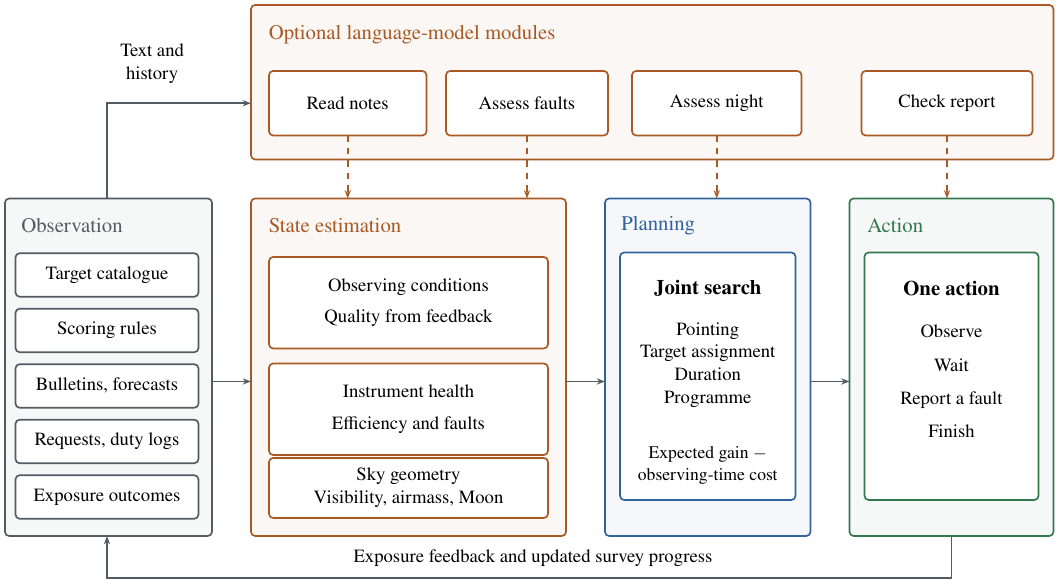}
\caption{Decision loop of the baseline model. Exposure feedback updates estimates of observing conditions, instrument health and sky geometry. A rule-based planner jointly selects pointing, target assignment, exposure duration and programme using expected gain and an observing-time cost. Optional language-model modules process duty logs, forecasts and quality history; dashed arrows indicate where their outputs modify the loop, with rule-based fallbacks. Baseline~1 uses a simpler exposure-selection planner with optional language-model advice.}
\label{fig:baseline}
\end{figure*}

Baseline~2 adds explicit estimation of observing conditions and instrument health and a joint search over pointing, assignment, duration and programme (Figure~\ref{fig:baseline}). Its planner balances expected gain against a cost for observing time. Optional language-model modules interpret operational text and forecasts and assess evidence for instrumental faults; their outputs modify parts of the numerical decision loop, with rule-based fallbacks. The two baselines therefore provide distinct combinations of state estimation, planning and model assistance. Their planning heuristics are described in Appendix~\ref{app:reference_details}; the common evaluation objective remains equation~\ref{eq:total}.

The final comparison uses the Python implementations of both baseline models published with the final leaderboard. Each has three scored runs on each held-out task, with the same score aggregation as the participant systems (Section~\ref{sec:evaluation}). Their scores are reported in Section~\ref{sec:reference_results}. We use Baseline~2 as the baseline model for the L3 reference campaign (Section~\ref{sec:localdiagnostic}).

\section{Evaluation and survey outcomes}
\label{sec:challenge}
We apply the common interface and objective of Section~\ref{sec:benchmark} to examine survey outcomes under fixed task conditions. The L3 reference campaign links individual decisions to target completion. The E--H evaluation compares submitted systems across four held-out campaigns and repeated runs. We then examine representative strategies and exploratory tests of operational responses and revised survey priorities.

\subsection{Evaluation sample and protocol}
\label{sec:evaluation}
AOH brought together an international field of teams from multiple countries. The online competition ran from 2026 October 5 to 7, following online training on October 2--3. Teams of one to three members developed and tested their observing systems on the public tasks, then submitted them through the competition platform. Final submissions were evaluated on the four held-out campaigns E--H using a common scoring procedure. The published final leaderboard contains 102 ranked teams. The participation and submission rules are available on the challenge website.\footnote{\url{https://create.gosim.org/survey26/rules}}

Award eligibility required language-model-driven technology in at least two stages, such as interpretation, planning, action selection or adaptation. Deterministic components and external model services were permitted. This eligibility rule shapes the evaluated sample.

Each submission was fixed to a source commit and container image, with the submitted system isolated from the held-out scenario products and scorer. The published computation budget was 900\,s of normalised time per run, with a 30-min elapsed-time cap. External inference services and external compute were permitted, so total resource use also depends on each system's external services. Model-service use and cost remain relevant properties of the comparison \citep{kapoor2025agents}. Task validity, outcome validity and transparent reporting are complementary requirements for agentic evaluation \citep{zhu2025agenticbenchmarks}. Appendix~\ref{app:clock} defines the computation accounting, and Appendix~\ref{app:evaluation_records} describes the execution and archival procedures.

Under the regular final protocol, each frozen submission is evaluated three times on each of E--H. The final score is the equally weighted mean of the four per-task means. Repeats measure variation of the submitted system on a fixed task, including stochastic decisions and service behaviour. Platform-attributable failures trigger replacement evaluations; the competition rules assign zero to team-attributable evaluation failures. Runs that reach a simulator termination code, including the elapsed-time limit, settle the observing history accumulated so far. The teams ranked second and 73rd were finalised using a different selection of valid runs. The results below use the published scores; Section~\ref{sec:repeatability} describes these exceptions and Appendix~\ref{app:execution_records} gives the full accounting.

\subsection{Survey outcomes in a reference campaign}
\label{sec:localdiagnostic}
We run the baseline model on local task L3 with language-model advice disabled. We analyse its complete archive to relate exposure choices and operational responses to retained target completion. The policy completes the 37-night calendar, spanning 2026 November 16, 00:30 UTC to December 22, 08:15 UTC in simulated observing time, with 957 exposures. Of 8700 input targets, 6474 reach $G_i\geq0.5$, including 434 of the 435 required targets: completion fractions of 74.4\% and 99.8\%, respectively. Both observing requests are met. The policy reports and repairs an instrumental fault eight days after its onset.

Figure~\ref{fig:timeline}(c) shows the events and information released during the campaign. The sharp efficiency loss around day~17 follows the first earthquake on December~3; an independent instrumental fault starts on December~7. Panel~(d) compares the executed programme labels with the maximum weighted response bonus at fixed pointing, assignments and duration. Before the first earthquake, 18 of 582 choices fall below this reference, compared with 127 of 210 between that event and fault repair. After repair, five of 165 do so, across a period that includes a second earthquake. The baseline uses saturated exposure feedback to estimate the prescribed quality band separately from response quality. Its default fallback and fit selection also depend on response quality (Appendix~\ref{app:reference_heuristics}). Instrument losses therefore bias band estimates within this feedback-based estimator. Appendix~\ref{app:reference_reconstruction} gives the event times, state estimates and fault reports.

Programme labels identify bands of the synthetic quality statistic in equation~\ref{eq:band}. On November~23, near full Moon, 525 of 558 target--exposure records occupy the high band, labelled DARK, under the prescribed lunar factor and site conditions. On December~9, near new Moon, all 146 records occupy the intermediate band, labelled BRIGHT; nine of the 16 baseline exposures instead select BACKUP. These examples test the configured band rule and its inference from exposure feedback. The post-hoc reference compares the resulting score multipliers for each executed exposure.

\begin{figure*}[ht!]
\centering

\includegraphics[width=0.985\textwidth]{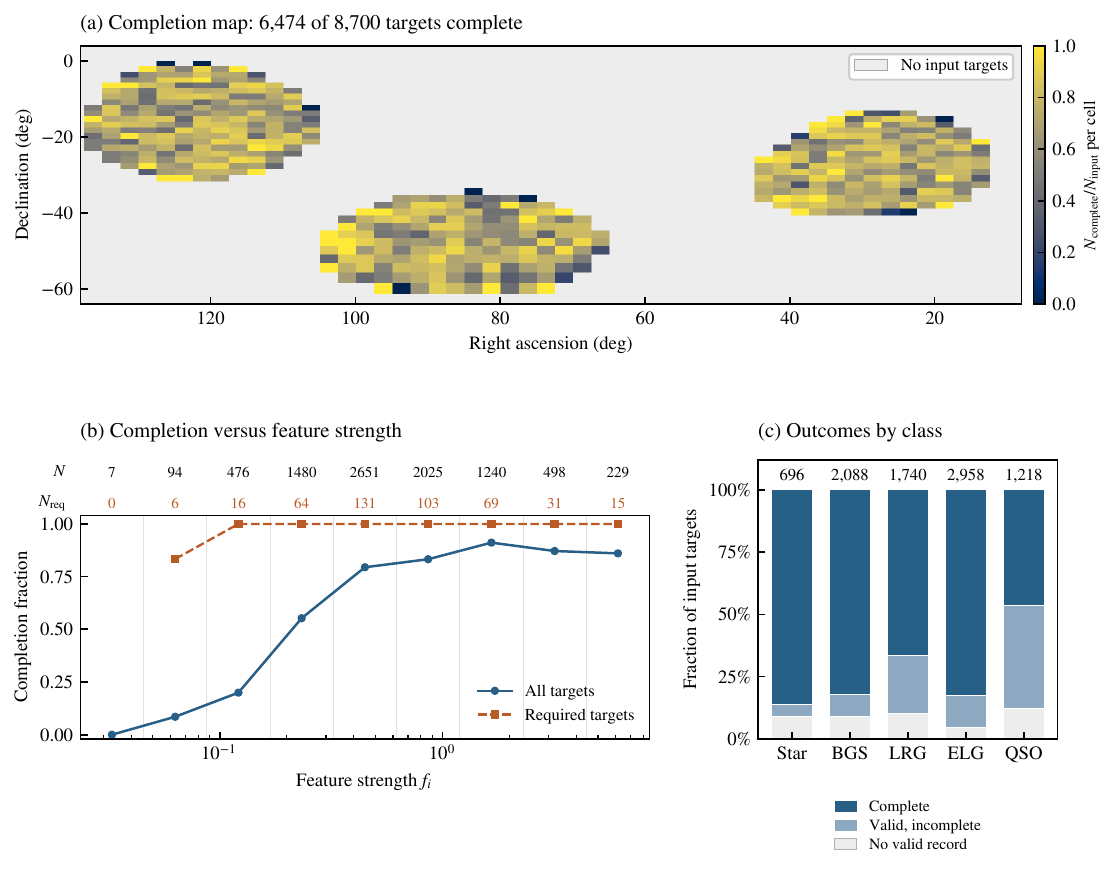}

\caption{Target coverage in the L3 reference run of Figure~\ref{fig:timeline} (baseline model, language-model advice disabled). (a) Completion fraction in equal-area cells of $2.5^\circ$ in right ascension by 0.025 in $\sin\delta$, drawn at their celestial-coordinate limits. Light-grey cells contain no input targets; dark blue marks populated cells with zero completion. Overall, 6\,474 of 8\,700 targets are complete. (b) Completion versus feature strength $f_i$, with total and required-target counts ($N$, $N_\mathrm{req}$) above each bin. At fixed observing quality, larger $f_i$ requires a shorter exposure to reach the threshold. (c) Fractions of each class with no surviving valid exposure record, with a valid record but incomplete, and complete; input counts appear above the bars. Among the 7\,983 targets with valid records, 11 have zero response in every retained exposure.}
\label{fig:coverage}\label{fig:completion}
\end{figure*}

Figure~\ref{fig:completion} shows target completeness as a function of position, class and feature strength. Completion generally increases with feature strength, as expected from the shorter exposure time needed at fixed observing quality. The modest decline in the highest bins coincides with an increasing fraction of low-weight stars. The highest bin contains 229 stars ($w_i=0.3$): 197 are complete, including all 15 required targets; 19 have no surviving observation and 13 remain below threshold.

\subsection{Performance across held-out campaigns}
\label{sec:final}
\label{sec:reference_results}
\begin{figure*}[ht!]
\centering
\includegraphics[width=0.77\textwidth]{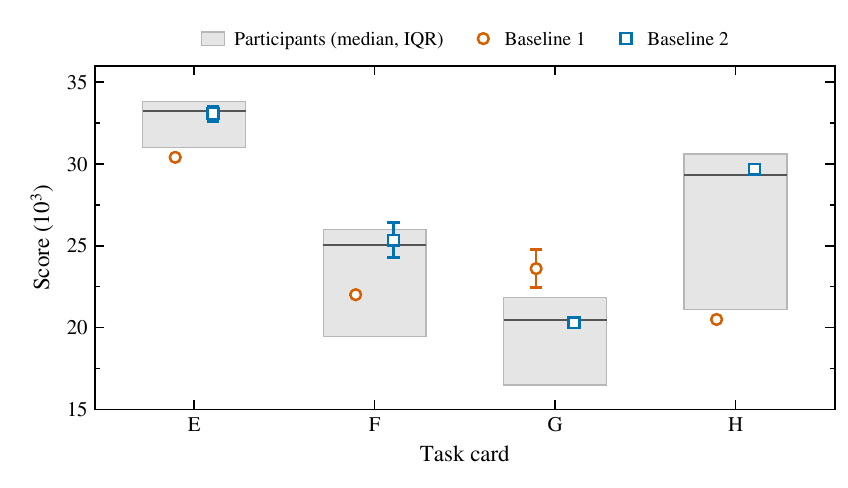}
\caption{Performance on the four held-out task cards. Shaded boxes span the 25th--75th percentiles of the 102 participant task means, and horizontal bars mark their medians (Weibull quartiles). Coloured symbols show the two baseline models, Baseline~1 and Baseline~2; error bars give the sample standard deviation of three runs. Symbols are offset horizontally for clarity.}
\label{fig:references}
\end{figure*}

The baseline models score 24\,134 for Baseline~1 and 27\,101 for Baseline~2 (Figure~\ref{fig:references}; Table~\ref{tab:main}).

The 102 ranked teams have a median final score of 27\,047 and an interquartile range of 22\,548--27\,864 (Figure~\ref{fig:results}a in Appendix~\ref{app:case_studies}). Seventy-four teams exceed the published Baseline~1 entry, and 46 exceed Baseline~2, which has the higher aggregate reference score. The highest-ranked system scores 33\,576 and leads on every task; the first three teams lie within 977 points and the first ten within 3\,045. Table~\ref{tab:main} gives the detailed scores.

Median team scores range from 20\,462 on G to 33\,251 on E (Figure~\ref{fig:results}b). The relative ordering of teams is broadly consistent across the four tasks, with tie-corrected Kendall's $W=0.86$. Each task combines a different catalogue, observing geometry, season and event history.

\subsection{Representative observing strategies}
\label{sec:allteams}
We describe representative observing strategies using the organisers' archived code-review report of 2026 October 9. The teams are identified by their published final ranks.

\begin{figure*}[ht!]
\centering
\includegraphics[width=0.96\textwidth]{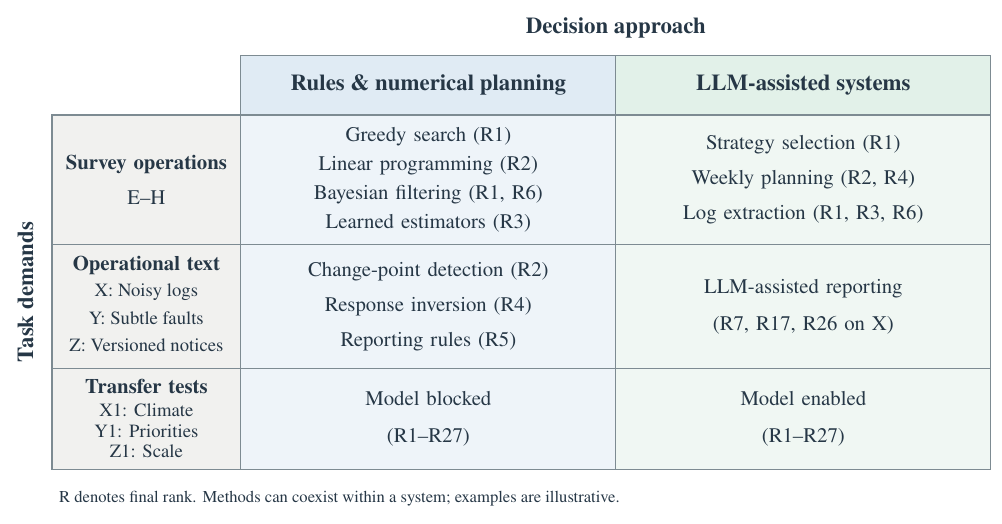}
\caption{Numerical and language-model components of selected AOH systems. Rank labels R1, R2, etc. refer to the final E--H leaderboard. The upper row shows observing methods described in the organisers' October 9 code-review report. The middle row shows fault-detection methods, model-assisted reporting gains measured for R7, R17 and R26 on X, and the notice-processing tasks Y and Z. The lower row gives the model-access settings for the same 27 selected systems in the transfer tests, using existing fallbacks when access is blocked. X1 changes climate, Y1 changes scientific priorities and scoring rules, and Z1 changes catalogue scale and operational disturbances. A system can use components from both columns. Sections~\ref{sec:allteams} and~\ref{sec:exploratory} describe the methods and results.}
\label{fig:allteams}
\alttext{A two-column, three-row matrix of numerical methods and language-model roles. Examples include greedy search, linear programming, Bayesian filtering, learned estimators, strategy selection, weekly planning, log extraction and fault reporting. Rows identify survey operations, operational-text tasks X/Y/Z and transfer tests X1/Y1/Z1. R labels denote final ranks, and methods can coexist within a system.}
\end{figure*}

Figure~\ref{fig:allteams} summarises representative methods and the task demands examined in the final and exploratory evaluations. Labels R1, R2, etc. identify the published final E--H ranks. The methods in its two columns can coexist within a system. All six leading entries use hand-written exposure planners, validate model outputs in code and fall back to rules. Their numerical components include greedy search (R1), a seasonal linear programme that sets observing-time prices (R2), Bayesian filtering (R1 and R6), and offline-trained condition and fault estimators (R3). Language models support strategy selection, weekly planning and extraction of facts from duty logs. Change-point detection (R2), per-exposure response inversion (R4) and reporting rules (R5) provide additional routes to fault diagnosis and recovery. Appendix~\ref{app:case_studies} gives the system-level descriptions; Section~\ref{sec:exploratory} reports the tests summarised in the lower rows of the figure.

\subsection{Run-to-run variability}
\label{sec:repeatability}
Repeat variation limits how finely the final ranking can be interpreted. In 2000 bootstrap resamplings (random seed 20261024), the three counted repeats are sampled with replacement independently for each team and task; the final mean scores are then ranked. This gives 90\% rank intervals spanning at most two positions for each top-ten team, but a median width of 20 positions for ranks 31--60 (Figure~\ref{fig:results}c). Across the 408 team--task groups, the median sample standard deviation is 307 points. Propagating the four task variances to each final mean gives a median standard error of 149 points across teams. The median gap between adjacent ranks is 92 points, and 42 teams lie between 26\,500 and 28\,000. Eighteen of 101 adjacent pairs differ by more than 1.96 combined standard errors. The middle of the ranking is therefore better interpreted as performance groups than as a precise ordering.

Two finalisations depart from the regular averaging protocol. The teams ranked second and 73rd were scored using the three highest-scoring valid runs on each task. Our summaries retain this published selection. Their bootstrap intervals describe resampling variability conditional on these selected runs. Selecting the highest observed outcomes can introduce selection effects; an analogous issue for model variants on Chatbot Arena is analysed by \citet{singh2025leaderboard}. Here the selected units are repeated runs of submitted systems. The candidate counts and selection rule are therefore part of the evaluation record (Appendix~\ref{app:execution_records}).

The intervals quantify run-to-run variability on the four fixed task realisations. Service conditions can vary between replacement attempts and can be shared by concurrently launched repeats. Of the 1\,200 counted runs for the 100 regularly finalised teams, 427 have a non-zero replacement counter; in 103 of their 400 task groups the three counted runs have no common execution-time overlap. These features can correlate or shift outcomes and limit interpretation of the nominal intervals. Detailed timing and termination records are summarised in Appendix~\ref{app:execution_records}.

\subsection{Exploratory stress and transfer tests}
\label{sec:exploratory}
Additional tests examine responses to operational text and changes in campaign conditions and scientific priorities (Figure~\ref{fig:allteams}). They use the frozen final versions of the 27 highest-ranked teams, with model calls routed to a single model and one run per tested configuration. Model-enabled runs are compared with runs that block model access and invoke each system's existing fallback. The X task also supplies a structured-summary comparison. Appendix~\ref{app:exploratory} gives the task definitions and detailed outcomes.

In X, a guide-camera failure is announced in a noisy, multilingual duty log. Across 25 paired submissions, the median score difference between model-enabled and model-blocked runs is approximately $+3\,400$. R7, R17 and R26 gain approximately 64\,000--72\,000 points with model access, correctly reporting 69--86 faults compared with 0--4 when access is blocked. The six highest-ranked systems score 26\,000--33\,000 with model access blocked; the severe efficiency loss makes the fault detectable through exposure feedback. Y uses subtle throughput changes and decoys, while Z adds engineering notices with revisions and references to earlier nights. The median score differences $S_{\mathrm{on}}-S_{\mathrm{off}}$ are approximately $+400$ on Y and $-100$ on Z. In these two tasks, most systems report few of the announced services.

The transfer tests change climate (X1), scientific priorities and scoring rules (Y1), or catalogue scale together with operational disturbances (Z1). Each task supplies its new parameters in the public configuration. For model-enabled runs, the Spearman rank correlations with the final evaluation are $\rho=0.85$, $0.38$ and $0.87$ for X1, Y1 and Z1, respectively. The largest changes occur on Y1, where large required-target penalties accompany the largest rank declines.

\section{Discussion}
\label{sec:discussion}

We discuss the implications of these results for survey planning and operational diagnostics, together with the requirements for adapting the simulation to a specific observatory.

\smallskip
\noindent \textit{Scientific priorities.} The final comparison shows that 46 of 102 systems improve on both baselines under the common survey objective. For the 27 selected systems, model-enabled transfer tests give rank correlations with the final evaluation of 0.38 for Y1, 0.85 for X1 and 0.87 for Z1. Y1 changes several scientific and scoring settings together, including the required-target fraction, the missed-target penalty and the false-report allowance. X1 changes climate, while Z1 changes catalogue scale and operational disturbances. A policy comparison for an observatory should therefore begin with its target priorities, depth and cadence requirements, which determine the value of each observation.

\smallskip
\noindent \textit{Survey diagnostics.} The L3 diagnostics illustrate the challenge priorities at target level. The baseline completes 74.4\% of all targets and 99.8\% of the required subset, while meeting both observing requests. Its completion maps and class fractions locate the targets with no retained observation or a response below threshold. The archived sequence also traces band-selection errors to the baseline's use of weighted response feedback. This finding concerns inference of the prescribed scoring bands. A facility study would compare inferred conditions with independent measurements of throughput, sky background and image quality, following the measured inputs used in exposure control \citep{schlafly2023operations}. On X, three systems correctly report more faults and achieve higher scores with model access. The large efficiency loss is also visible in exposure feedback, and the six leading systems retain high scores with model access blocked. These simulated results support further comparisons of diagnostic policies using both measured exposure outcomes and operational notices.

\smallskip
\noindent \textit{Completion criteria.} The completion rule and time accounting determine exposure allocation. For one target at fixed observing quality, with all exposure records valid, the largest-single-exposure rule gives at least as much retained raw response when a fixed integration time is concentrated in one visit as when it is divided among shorter visits, provided that the single visit satisfies the duration and visibility limits. Zero transition overhead independently removes the time cost of slews, reassignment and readout. These assumptions shape the value of revisiting a field. DESI splits long observations to manage cosmic rays and changing geometry, and combines the spectra in its data processing \citep{schlafly2023operations,guy2023pipeline}. A facility model would use accumulated depth and measured overheads, together with the relevant exposure-splitting constraints. Imaging applications would also include filter choice and cadence.

\smallskip
\noindent \textit{Observatory applications.} A facility comparison would combine target responses and site histories calibrated against exposure and monitoring data with an explicit model of feasible actions. OpSim models survey execution \citep{delgado2014opsim}, while MAF evaluates observing histories through science metrics \citep{jones2014maf}. For a fibre spectrograph, feasible assignments depend on patrol regions and collision constraints \citep{desi2022instrument,silber2023focalplane,smith2019fiberassignment}. Calibration exposures, measured slew and readout costs, and safety requirements would enter the observing model; Galactic extinction and crowding would enter target selection. Policies could then be compared on the same catalogue and condition histories using completeness, achieved depth, cadence and recovery time. Recording actions and outcomes would allow changes in survey performance to be traced to observing decisions. The released simulator and reference strategies provide a starting point for these facility-specific comparisons.

\section{Conclusions}
\label{sec:conclusions}
We have presented AOH, a simulation of survey observations with target assignment, exposure planning, time-dependent observing requests and recovery from operational interruptions. Among 102 systems evaluated on four held-out campaigns, 46 score above both baseline models. The leading system scores 33\,576 and has the highest mean score on each campaign. The six leading systems combine numerical exposure planners with language models for planning and interpretation of operational records.

The L3 reconstruction follows retained target completion through changing conditions and diagnoses errors in the baseline's inference of synthetic quality bands. This provides a diagnostic example alongside the comparison across held-out campaigns. Exploratory tests show more correct fault reports with model access in several systems and substantial changes in ranking under revised scientific priorities. A facility application requires a science objective, calibrated exposure and operating models, and constraints appropriate to its instrument. The released simulation and reference strategies support comparisons of observing decisions under these specified models.

\begin{acknowledgments}
AOH is supported by the GOSIM open-source community.

Generative-AI assistance was used to organise and edit this draft and to assist reference and implementation checks. 

\end{acknowledgments}

\section*{Author contributions}
Haosong Wang contributed to manuscript writing, simulation development, and challenge engineering and organisation. Ting Tan and Huanyuan Shan co-led the challenge and contributed to manuscript writing and simulation development. Yao Li contributed to manuscript writing and challenge engineering and organisation. Dezi Liu, Xiangkun Liu, Yifei Luo, Wei Wang, Zonghuan Wu and Zhongxu Zhai contributed to challenge organisation throughout the project, provided scientific advice and reviewed the manuscript.

\FloatBarrier
\section*{Data Availability}
The AOH simulator, scoring code, reference implementations and documentation are available at \url{https://github.com/gosimfoundation/hackathon-survey26}. The repository also contains public replay data, including target catalogues, observing conditions and action histories. Released task bundles are listed at \url{https://create.gosim.org/survey26/platform/resources}, and the final leaderboard and campaign replays are available at \url{https://create.gosim.org/survey26/platform/leaderboard/final-hidden}. Our score analysis uses the official per-run export of 2026 October 11. Teams can access their full execution logs through the result repositories linked to their platform accounts.
\clearpage
\appendix
\restartappendixnumbering
\section{Simulation geometry and response models}
\label{app:conventions}
Table~\ref{tab:simulation} summarises the settings shared by the final E--H tasks. We first describe observing geometry and exposure integration, then the generation of catalogues, weather and events.

\begin{table*}[ht!]
\centering\small
\caption{Common simulation parameters for the final E--H tasks.}
\label{tab:simulation}
\begin{tabular}{>{\raggedright\arraybackslash}p{4.0cm}p{2.2cm}p{2.4cm}>{\raggedright\arraybackslash}p{7.1cm}}
\toprule
Quantity & Symbol & Value & Role in the simulation\\
\midrule
\multicolumn{4}{l}{\textit{Observing geometry and time}}\\
Solar-altitude boundary (deg) & --- & $-18$ & Defines astronomical darkness.\\
Minimum target altitude (deg) & --- & 30 & Enforced throughout the actual exposure.\\
Weather interval (s) & --- & 900 & Discretises conditions and usable night boundaries.\\
Requested exposure duration (s) & --- & 60--3600 & Integer seconds; actual duration is limited by night and campaign boundaries.\\
Assignment-cell gap (deg) & $g$ & 0 & Contiguous square cells; at most one target per cell per exposure.\\
Exposure-transition overhead (s) & --- & 0 & The next decision follows the end of the exposure.\\
Integration interval bound (s) & $\Delta t_p$ & $\leq120$ & Midpoint quadrature, with additional splits at relevant boundaries.\\
\midrule
\multicolumn{4}{l}{\textit{Exposure response}}\\
Quality normalisation & $q_0$ & 0.68 & Sets the dimensionless quality scale.\\
Airmass exponent & $\beta$ & 0.60 & Controls the response dependence on airmass.\\
Lunar penalty amplitude & $P$ & 0.75 & Sets the amplitude of the prescribed lunar factor.\\
Lunar-altitude exponent & $\gamma$ & 1 & Controls the lunar-altitude dependence.\\
Lunar angular scale (deg) & $\theta$ & 35 & Sets the decay scale with target--Moon separation.\\
\midrule
\multicolumn{4}{l}{\textit{Observing-programme assignment}}\\
DARK band & $b_{ie}$ & $b_{ie}\geq0.65$ & Matching programme gives multiplier $m_{ie}=1.20$.\\
BRIGHT band & $b_{ie}$ & $0.40\leq b_{ie}<0.65$ & Matching programme gives multiplier $m_{ie}=1.12$.\\
BACKUP band & $b_{ie}$ & $b_{ie}<0.40$ & Matching programme gives multiplier $m_{ie}=1.06$.\\
Mismatched programme & $m_{ie}$ & 1.00 & Retains the unenhanced science-weighted response.\\
\bottomrule
\end{tabular}
\tablecomments{Entries describe the released task settings and shared numerical conventions. The response quantities are defined in equations~\ref{eq:quality}, \ref{eq:response} and \ref{eq:band}; the lunar factor is defined in Appendix~\ref{app:conventions}. Site, calendar and catalogue size vary by task (Table~\ref{tab:cards}), as do assignment area and the response scales $f_0,T_0$ (Table~\ref{tab:operations}). Weather-generation coefficients belong to each scenario's adopted parameter set.}
\end{table*}

\subsection{Observing geometry and calendar}
\label{app:geometry}
Target altitude is calculated from celestial coordinates and local sidereal time \citep{smart1977spherical}. Solar and lunar positions use the simulator's analytic approximations. Model twilight crossings are refined with a one-second stopping tolerance; dusk is rounded forward and dawn backward to an epoch-aligned 900-s grid. This retains complete weather intervals within astronomical darkness and gives a date-dependent night length. Conditions are stored as constant values within each interval. Exposure requests use integer-second durations. An exposure starting in a calendar gap ends at the earlier of its requested end and the campaign end, with gap time contributing zero response.

For assignment-cell side $d$ and gap $g$, the grid pitch is $d+g$ and its outer side is $n(d+g)$ in tangent-plane units; the final E--H configurations use $g=0$. The gnomonic projection follows \citet{calabretta2002projection}, with axes along increasing local altitude and azimuth at exposure start and a scale factor of $180/\pi$. Cell membership is evaluated at exposure start and held fixed throughout the exposure. The projection applies to angular separations smaller than $90^\circ$. A geometrically eligible assignment requires the actual field centre to lie in the altitude range $[0^\circ,90^\circ]$, the target to occupy its assigned cell at exposure start, and its altitude to remain at or above $30^\circ$ throughout the actual exposure. The $30^\circ$ altitude limit corresponds to approximately $X\lesssim2$ in the adopted airmass model.

\subsection{Exposure quality and numerical integration}
\label{app:response_integration}
We adopt the functional form of \citet{kasten1989airmass} for the airmass,
\[
A(z)=\left[\cos(\pi z/180)+0.50572(96.07995-z)^{-1.6364}\right]^{-1},
\qquad X(z)=\frac{A(z)}{A(0)},
\]
where $z$ is the numerical value of the geometric zenith angle in degrees. The published approximation uses apparent zenith angle; inserting the geometric angle is a simulator convention. The division by $A(0)$ normalises the airmass at the zenith.

The lunar factor is
\begin{equation}
L_i=1-P I\,[\sin(\max(0,h_{\mathrm{M}}))]^\gamma
\exp(-\rho_i/\theta),\label{eq:lunar}
\end{equation}
where $h_{\mathrm{M}}$ is lunar altitude, $\rho_i$ is target--Moon angular separation, and $I=(1-\cos\psi)/2$ is the adopted illuminated fraction from Sun--Moon elongation $\psi$. The E--H values of $P$, $\gamma$, and $\theta$ are listed in Table~\ref{tab:simulation}; $\rho_i$ and $\theta$ use the same angular units. Trigonometric arguments are converted consistently to radians in the implementation. The factor reduces the response as the illuminated Moon rises or approaches the target. Its separable form assigns distinct controls to loss amplitude, altitude weighting and angular decay, with $1-P\leq L_i\leq1$.

For programme selection, the condition-based quality is
\begin{equation}
b_{ie}=\frac{1}{q_0T_e}
\sum_{p\in\mathcal W_e}\Delta t_p D^{\mathrm{site}}_p
\frac{\tau_p\kappa_pL_i(t_p)}
{s_pX_i(t_p)^\beta},
\label{eq:band}
\end{equation}
Here $\tau_p$, $\kappa_p$ and $s_p$ are the stored site factors, including site-wide event effects, and $D^{\mathrm{site}}_p$ is one for an open site interval and zero for closed intervals or calendar gaps. This prescribed statistic assigns the synthetic quality band used by the scoring rule. The DARK, BRIGHT and BACKUP thresholds and matching multipliers are listed in Table~\ref{tab:simulation}; an unmatched programme has $m_{ie}=1$.

Equations~\ref{eq:quality} and \ref{eq:band} use separate midpoint partitions. The full-response partition $\mathcal P_e$ splits at weather boundaries and the start and end times of directional events. The programme partition $\mathcal W_e$ splits at weather boundaries. Both subdivide each segment into pieces of at most 120\,s. Directional-sector membership is evaluated at each piece midpoint; a target crossing a sector within a piece is therefore represented by midpoint sampling. Closed pieces contribute zero numerator with their duration retained in $T_e$. The target altitude constraint is checked throughout the complete actual exposure before scoring.

Site-wide event factors modify the stored conditions before clipping to their configured bounds. Directional effects are evaluated from target altitude and azimuth at the integration midpoints; closures and obstructions give zero response there. Earthquake recovery is stored in a separate effect table and can extend beyond the initial event interval.

\subsection{Footprints and target catalogues}
\label{app:generators}
For region $j$, we choose a centre and place $K$ boundary vertices at equally spaced position angles $\vartheta_k=2\pi k/K$, where $k=0,\ldots,K-1$. The angular distance of each vertex from the centre is
\begin{equation}
R_j(\vartheta_k)=b_j\left[1+\sum_{m=1}^{M}
a_{jm}\cos(m\vartheta_k+\psi_{jm})\right].
\label{eq:footprint}
\end{equation}
Here $b_j$ sets the angular scale, $a_{jm}$ and $\psi_{jm}$ set the amplitude and phase of harmonic $m$, and $M$ is the number of harmonics. The amplitudes and phases are drawn using a fixed random seed, with amplitude bounds that keep all radii positive. With zero amplitudes, the vertices lie at equal angular distances from the centre; the harmonic terms introduce boundary variations, with larger $m$ describing finer angular structure. We join adjacent vertices by great-circle arcs and adjust $b_j$ until the spherical polygon matches its assigned area within the prescribed tolerance.

For each region $j$, we sum the signed solid angles of triangles formed by its centre and adjacent vertices \citep{vanoosterom1983solidangle,jacobson2013winding}. We take the absolute sum, replace areas above $2\pi$ by their complement to $4\pi$, and convert steradians to square degrees to obtain $\Omega_j$. Starting from the angular radius of a spherical cap with the assigned area $\Omega_j^{\mathrm{target}}$, we update the radial scale as $b_j\leftarrow b_j\sqrt{\Omega_j^{\mathrm{target}}/\Omega_j}$. We reconstruct the vertices and recompute the spherical area after each update, stopping when the relative area error falls below the prescribed component tolerance. The area ratio is dimensionless, with both areas expressed in the same units.

For a diffuse proposal, we select a footprint component in proportion to its area and draw a position in its enclosing disc, with angular radius $r=R_{\max}\sqrt{U}$ and position angle $2\pi V$, where $U$ and $V$ are independent uniform variates on $[0,1]$. This radial law samples area uniformly in the plane of angular offsets. Clustered proposals use independent Gaussian east--north offsets around centres sampled within the footprint. We map each proposal onto the sphere and retain positions that lie within the footprint and have an observing window during the season. The spherical mapping and rejection cuts determine the final surface density and mixture fractions.

For a catalogue of size $N$, class counts follow the specified fractions, with rounding corrections that preserve $N$; class labels are randomly assigned to the accepted targets. For class $c$, we draw $\ln f_i\sim\mathcal N(\ln\widetilde f_c,\sigma_{\ln f,c}^{2})$, where $\widetilde f_c$ is the median feature strength and $\sigma_{\ln f,c}$ its logarithmic scatter \citep{nist2012handbook}. All targets of class $c$ receive the science weight $w_c$. The generator selects $\operatorname{round}(r_{\mathrm{req}}N)$ distinct target indices with equal probability as required targets, where $r_{\mathrm{req}}$ is the specified fraction.

\subsection{Weather and operational events}
\label{app:weather}\label{app:weather_comparison}
For weather variable $a$, let $\mathcal C_a(x)=\max[\ell_a,\min(u_a,x)]$ clip to its configured bounds. The seasonal mean is $\mu_{a,n}=\nu_a+\epsilon_a A_a\sin\theta_n$, with $\theta_n=2\pi[d(n)-d_\star]/365.2425$. Here $d(n)$ is the local night's day of year, $d_\star$ is the phase day, $\nu_a$ is the nominal value, $A_a$ is the amplitude, and $\epsilon_a$ is $+1$ for seeing and $-1$ for transparency and sky quality. The nightly state $Y_{a,n}$ and intra-night value $x_{a,n,k}$ evolve as
\begin{align}
 Y_{a,n}&=\mathcal C_a\bigl[\mu_{a,n}+\phi_{\mathrm{N}}(Y_{a,n-1}-\mu_{a,n})
 +\sigma_{a,\mathrm{N}}\sqrt{1-\phi_{\mathrm{N}}^2}\,\xi_{a,n}\bigr],\label{eq:nightweather}\\
 x_{a,n,k}&=\mathcal C_a\bigl[Y_{a,n}+\phi_{\mathrm{S}}(x_{a,n,k-1}-Y_{a,n})
 +\sigma_{a,\mathrm{S}}\sqrt{1-\phi_{\mathrm{S}}^2}\,\zeta_{a,n,k}\bigr].\label{eq:slotweather}
\end{align}
Here $n=1,\ldots,N_{\mathrm{night}}$ indexes nights and $k=1,\ldots,K_n$ indexes intervals within night $n$. The coefficients $\phi_{\mathrm{N,S}}$ set the persistence, and $\sigma_{a,\mathrm{N,S}}$ set the fluctuation scales. The variables $\xi$ and $\zeta$ are independent standard normals, so the innovation standard deviations are $\sigma_{a,\mathrm{N,S}}\sqrt{1-\phi_{\mathrm{N,S}}^2}$. This scaling makes the configured fluctuation scale the stationary standard deviation of the corresponding constant-mean, unclipped autoregression. We initialise the states with $Y_{a,0}=\nu_a$ and $x_{a,n,0}=Y_{a,n}$. The nightly update centres the previous state on the current night's seasonal mean. Seasonal evolution, clipping and initialisation determine the resulting variances and correlations. This construction gives bounded weather histories with variations within and between nights. Background instrument-efficiency fluctuations follow a uniform distribution. Each task stores the resulting conditions, including event effects, for use by all evaluated policies.

For E--H, the nominal seeing is $1.10\,\mathrm{arcsec}$, the seasonal amplitude is $0.15\,\mathrm{arcsec}$, and the fluctuation scales are $\sigma_{s,\mathrm N}=0.16\,\mathrm{arcsec}$ and $\sigma_{s,\mathrm S}=0.07\,\mathrm{arcsec}$. The persistence coefficients are $\phi_{\mathrm N}=0.82$ and $\phi_{\mathrm S}=0.88$, with conditions stored every 900\,s. The within-night seeing innovation therefore has standard deviation $0.07\sqrt{1-0.88^2}\simeq0.033\,\mathrm{arcsec}$ per interval. Site-wide event factors and closures modify these background conditions.

The availability chain is initialised in the open state before the first interval of each night, and a transition is drawn for each interval. The open-to-closed probability is $p_{\mathrm{close}}(n)=p_0+p_{\mathrm{seas}}\max(0,\sin\theta_n)$, where $p_0$ and $p_{\mathrm{seas}}$ are the baseline and seasonal amplitudes. The closed-to-open probability is $p_{\mathrm{reopen}}$.

The earthquake efficiency loss is
\begin{equation}
 D_{j,n}=\min\!\left[D_{\max},e^{\gamma_{\mathrm q}(M_j-M_\star)}\right]
 e^{-(n-n_j)/\tau_{\mathrm{rec}}},\quad n\geq n_j,
 \label{eq:quake}
\end{equation}
where $M_j$ is the event magnitude parameter, $M_\star$ the reference magnitude, $\gamma_{\mathrm q}$ the magnitude-scaling coefficient, $D_{\max}$ the initial loss cap, $n_j$ the onset night and $\tau_{\mathrm{rec}}$ the recovery time in nights. The cap bounds the initial loss, and $\tau_{\mathrm{rec}}$ sets the number of nights for an e-fold decrease. These parameters control the severity and duration of reduced efficiency for testing adaptation. The efficiency multiplier is $1-D_{j,n}$; configured coefficients optionally also scale seeing, transparency and sky quality. Effects begin at the event timestamp and continue while their amplitude reaches or exceeds the activity threshold. All coefficients belong to the versioned scenario configuration.

Standard event bulletins use eight compass sectors or a site-wide designation; the initial bulletin also identifies terrain sectors. The reference forecast cadence and horizon are seven days, covering weather and launch events by night. Earthquake effects appear in current bulletins, while exposure feedback and duty logs provide information about instrument performance. Request generation screens targets for visibility and exposure requirements at nominal quality unity. This identifies candidate observing opportunities; realised conditions and assignments determine whether the issued request is achieved.

\section{Task configurations and decision protocol}
\label{app:benchmark_details}
\subsection{Complete task configurations}
\label{app:task_details}
Table~\ref{tab:operations} gives the assignment, exposure and operational settings for the 20 tasks. Table~\ref{tab:task_design} gives their observing campaigns, target catalogues and evaluation roles.

\begin{table*}[ht!]
\centering\footnotesize\setlength{\tabcolsep}{5pt}
\caption{Assignment, exposure and operational settings of the 20 tasks.}
\label{tab:operations}
\begin{tabular}{lrrrrrrl}
\toprule
Card & $N_{\mathrm{cell}}$ & $A_{\mathrm{grid}}$ & $f_0$ & $T_0$ (s) & Requests & Faults & Stress\\
\midrule
L1 & 16 & 6.4 & 0.5 & 900 & 2 & 1 & no\\
L2 & 16 & 6.4 & 0.5 & 900 & 2 & 1 & no\\
L3 & 16 & 6.4 & 0.5 & 900 & 2 & 1 & yes\\
L4 & 16 & 6.4 & 0.5 & 900 & 2 & 1 & yes\\
\midrule
$\alpha$ & 16 & 6.4 & 0.5 & 900 & 2 & 1 & no\\
$\beta$ & 16 & 6.4 & 0.5 & 900 & 2 & 1 & no\\
$\gamma$ & 16 & 6.4 & 0.5 & 900 & 2 & 1 & yes\\
$\delta$ & 16 & 6.4 & 0.5 & 900 & 2 & 1 & yes\\
\midrule
A & 16 & 6.4 & 0.5 & 900 & 6 & 1 & no\\
B & 25 & 1.6 & 0.8 & 1000 & 6 & 1 & no\\
C & 9 & 6.3 & 0.5 & 900 & 6 & 1 & no\\
D & 100 & 6.0 & 0.5 & 900 & 6 & 3 & yes\\
\midrule
A1 & 16 & 6.4 & 0.5 & 900 & 11 & 104 & no\\
B1 & 25 & 1.6 & 0.8 & 1000 & 12 & 131 & no\\
C1 & 9 & 6.3 & 0.5 & 900 & 12 & 54 & no\\
D1 & 100 & 6.0 & 0.5 & 900 & 14 & 267 & yes\\
\midrule
E & 100 & 6.0 & 0.5 & 900 & 6 & 46 & no\\
F & 9 & 6.3 & 0.5 & 900 & 6 & 42 & no\\
G & 25 & 1.6 & 0.8 & 1000 & 6 & 32 & no\\
H & 16 & 6.4 & 0.5 & 900 & 6 & 29 & yes\\
\bottomrule
\end{tabular}
\tablecomments{Values are from the archived task products listed in Table~\ref{tab:cards}. $A_{\mathrm{grid}}$ is the total assignment area in deg$^2$ of tangent-plane coordinates. Requests counts observing requests with a completion reward; A1--D1 and E--H also provide free-text duty logs. Faults counts reportable instrumental faults. Stress marks cards with pointing offsets and the loss of earlier exposure records enabled. All cards use $q_0=0.68$.}
\end{table*}

The five task families use the information-access rules described in Section~\ref{sec:tasks}. Figure~\ref{fig:catalogues} shows the sky distributions of the final E--H catalogues.

Table~\ref{tab:populations_efgh} gives the target populations in E--H. The four catalogues share the same prescribed class fractions and science weights; integer target counts introduce small rounding differences in the realised fractions. Feature strengths are independently drawn for each catalogue, producing the empirical medians listed in the table.

\label{app:extra}

\begin{table*}[ht!]
\centering
\caption{{Synthetic target populations in the final test catalogues E--H.}}
\label{tab:populations_efgh}
{\begin{tabular}{lrrrrrr}
\toprule
Class & Prescribed fraction (\%) & $w_i$ & Median $f_i$: E & F & G & H\\
\midrule
Star & 8 & 0.30 & 4.012 & 4.008 & 3.958 & 3.996\\
BGS & 24 & 0.45 & 1.256 & 1.257 & 1.250 & 1.260\\
LRG & 20 & 1.00 & 0.349 & 0.350 & 0.349 & 0.348\\
ELG & 34 & 1.00 & 0.552 & 0.552 & 0.549 & 0.551\\
QSO & 14 & 1.70 & 0.184 & 0.181 & 0.179 & 0.180\\
\midrule
Targets & & & 30\,000 & 28\,351 & 35\,000 & 30\,000\\
Required & & & 1\,500 & 1\,418 & 1\,750 & 1\,500\\
\bottomrule
\end{tabular}}
\tablecomments{{Computed from the registered final scenario bundles. The four cards use the same prescribed class fractions and science weights; the final four columns give empirical medians of the dimensionless feature strength $f_i$, rounded to three decimal places.}}
\end{table*}

\subsection{Decision process and feedback}
\label{app:decision}
The decision system maintains its internal state throughout a campaign. At decision epochs $t_k$, the task can be described as a partially observable Markov decision process \citep{kaelbling1998pomdp}. The campaign has a fixed end time, while the number and spacing of decisions depend on the chosen actions. The state includes the clock, task inputs, complete observation record, telescope and instrument status, request and report accounting, and the unreleased weather and event schedule. The observation $o_k$ contains the information available to the system at $t_k$. A policy maps its observation--action history to an action
\[
a_k\in\{\textsc{observe}(p,\mathbf c,\pi,\Delta t),\;
\textsc{wait}(\Delta t),\;\textsc{report},\;\textsc{stop}\}.
\]
Here $p=(\mathrm{alt},\mathrm{az})$ specifies the pointing, $\mathbf c$ assigns at most one target to each of the task's $n^2$ cells, $\pi$ is the observing programme, and $\Delta t$ is the requested duration. The transition applies the action, advances the clock by its actual duration and updates the observation record and operational state. The episode ends at campaign termination or \textsc{stop}, with the settled score $S$ of equation~\ref{eq:total} as its terminal return. Including the complete observation and settlement state retains the dependence of completeness, uniformity and request outcomes on earlier actions.

At each decision point the interface provides the current time, latest bulletin and forecast, newly issued messages, active-request progress, the sum of best target scores, the previous action result and the remaining computation budget. A report returns its diagnostic outcome and reward or penalty. Ordinary exposure feedback provides geometrically accepted target identities and rounded programme-weighted scores; the raw factors used by the scorer remain internal. State updates following data loss communicate the surviving target progress. Accepted actions are associated with their decision points and retained together with their outcomes.

\subsection{Uniformity and operational settlement}
\label{app:settlement}
In the adopted benchmark setting, a single missing required target costs 50, compared with a largest individual target contribution of $1.70\times1.20=2.04$, and the uniformity penalty is bounded by 200. These scales make required-target completion valuable relative to the marginal contribution of an individual target; the total target-utility term still accumulates over the full catalogue.

For each populated 10-degree right-ascension band $k$, let $r_k$ be the fraction of catalogue targets with $G_i\geq0.5$. Across the $K$ populated bands, Jain's index is
\[
J=\frac{(\sum_{k=1}^{K}r_k)^2}{K\sum_{k=1}^{K}r_k^2}.
\]
We set $J=0$ when all completion fractions vanish; equal non-zero fractions give $J=1$. Empty catalogue bands are excluded. Thus the statistic compares completion fractions across populated bands without weighting by their target counts. The uniformity penalty is $200(1-J)$. Two-dimensional completion maps, such as Figure~\ref{fig:completion}a, complement this right-ascension statistic by resolving the angular structure of the selected sample.

For each observing request, the scorer takes the largest eligible raw response of each requested target using valid exposures wholly contained in the issue-to-deadline interval, including its endpoints. It counts targets reaching the request's threshold and awards the specified reward if the required count is reached. The request contribution is recomputed after any supporting data loss. Fault reports are processed against the active reportable fault and the false-report allowance. Correct reports reset that allowance; up to 32 consecutive report actions are permitted. All report rewards and penalties are included in $R_{\mathrm{rep}}$.

A data-loss trigger is processed at the first subsequent decision point, after completion of any exposure in progress. For $N$ completed exposures and configured fractional endpoints $a,b$, the invalidated action-index range is $[\lfloor aN\rfloor,\lfloor bN\rfloor)$. The implementation preserves the original ledger entries and updates their validity flags.

\subsection{Reference-policy heuristics}
\label{app:reference_heuristics}\label{app:reference_details}
The released Python implementation of Baseline~1 ranks visible targets by science value, required status and remaining visibility, then searches fields around selected anchor targets. After selecting a pointing and assignment, its duration search uses weighted increases in squared response estimates and a bonus of 60 for crossing the required-target completion threshold. Among candidate durations, it can prefer one that reaches a needed request-target threshold if its heuristic gain rate is at least 90\% of the best rate. The squared-response gain increases the relative value of larger predicted responses, while the threshold bonus favours completing required targets. These quantities rank candidate actions; the settled score uses equation~\ref{eq:total}, including the required-target penalty of 50. For programme selection, Baseline~1 scales its lunar and airmass model by an empirical response scale divided by a fixed reference efficiency of 0.95. The scale is inferred from exposure feedback after dividing out target strength, duration and modelled lunar and airmass factors. It therefore retains changes in instrument efficiency. The scorer's programme-band quality in equation~\ref{eq:band} uses atmospheric factors and target geometry.

Baseline~2 separately estimates the response scale used for exposure planning and the quality scale used for programme selection. Saturated target responses constrain programme matching through their observed score multipliers. With fewer than four such samples in the preceding two hours, the default implementation falls back to the response scale divided by 0.95; otherwise, equally consistent band fits are selected by proximity to that same estimate. Band inference consequently retains a dependence on instrument efficiency when feedback is sparse or ambiguous. The ratio of the two inferred scales supplies an instrument-efficiency proxy; the policy combines changes in this proxy with earthquake notices and the false-report allowance when deciding whether to report a fault. Its joint planner searches pointing, target assignment, duration and programme using incremental target value and an explicit cost of observing time. This cost balances the gain from the current exposure against time available for subsequent fields. Optional model stages provide night-level advice and fault assessment, with a further check before costly reports; these stages are disabled in the local L3 diagnostic run. The released planner also adjusts its search effort to the remaining computation budget. The archived actions preserve the diagnosed trajectory; a fresh policy execution also depends on measured computation costs. Source versions, settings and model outputs, when used, are needed to reproduce the evaluated configuration.

\section{Competition, execution and reconstruction records}
\label{app:evaluation_records}
\subsection{Competition sample and public rankings}
\label{app:competition_records}

The final leaderboard publishes 102 ranked participant entries and two baseline models. Its frozen data export, generated on 2026 October 11 at 10:37:50 UTC, provides run identifiers, per-run scores, task means, totals and execution metadata. The final-score analyses use this export.

Public resources, rules and the final leaderboard are linked in Data Availability.

\begin{figure*}[ht!]
\centering
\includegraphics[width=0.985\textwidth]{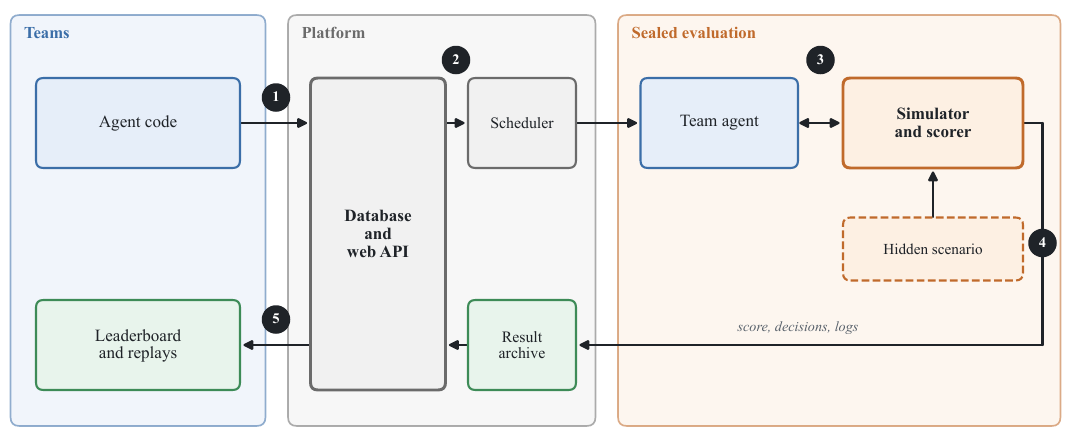}
\caption{Evaluation platform. Teams submit source versions, the scheduler queues isolated evaluations, and each submitted system exchanges decisions with the simulator, which accesses the held-out scenario. The platform archives the score, decision record and logs and publishes the score and replay. Colours distinguish the team side (blue), organiser platform (grey), isolated evaluation (orange) and outputs (green).}
\label{fig:platform}
\end{figure*}

\begin{figure*}[ht!]
\centering
\includegraphics[width=0.985\textwidth]{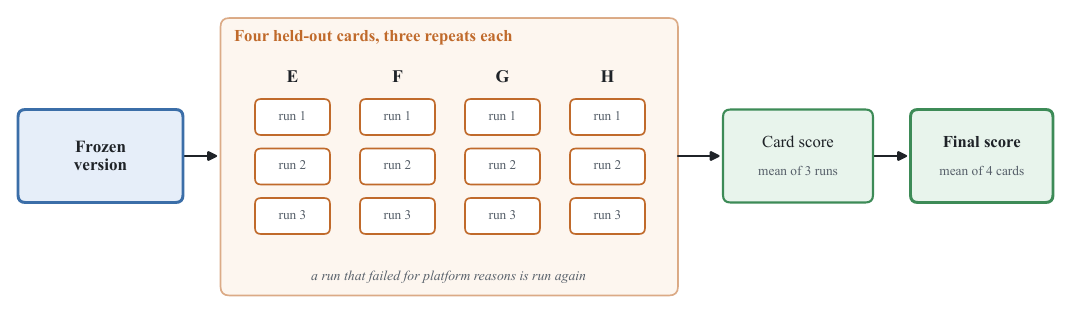}
\caption{Regular held-out evaluation protocol. Each frozen submission is evaluated three times on each of four tasks, with platform-attributable failures replaced. The final score is the mean of the four three-run means. Exceptions and the timing of counted replacement runs are described in Appendix~\ref{app:execution_records}.}
\label{fig:protocol}
\end{figure*}

\subsection{Compute accounting}
\label{app:clock}
For decision window $j$, let $W_j$ be elapsed time, $\Delta C_j$ the measured CPU-counter increment, and $a_j$ the calibrated duration ratio relative to a reference machine. The default metered implementation charges
\begin{equation}
 t_{\mathrm{charged}}=\sum_j\frac{\min\{W_j,\max(0,\Delta C_j)\}}{a_j}.\label{eq:clock}
\end{equation}
The calibration uses a rolling median duration for a fixed workload, with the submitted process paused during recalibration. A larger $a_j$ corresponds to a slower machine. If CPU metering becomes unavailable, the affected and subsequent windows use $W_j$ as their CPU surrogate. The elapsed-time cap begins at the first decision window and includes subsequent between-turn work. The public setting is 900\,s charged time and 1800\,s elapsed time; configuration and environment overrides belong in the archived execution record.

\subsection{Execution, termination and exceptional finalisations}
\label{app:execution_records}

Figure~\ref{fig:platform} outlines the evaluation platform. Each submission is fixed to a source commit and container image. The submitted system and the simulator run in separate jobs and communicate through the interface in Section~\ref{sec:interface}. The simulator accesses the held-out scenario products and scorer. At termination, the platform publishes the settled score and archives the decision trace, exposure ledger and logs in a team-specific result repository.

Each task runs in an isolated Linux container with two CPU cores and 2\,GB of memory, subject to the computation accounting in Appendix~\ref{app:clock}. The initial three repeats for a task are launched concurrently; counted replacements may start later. Platform-attributable failures trigger replacement evaluations under the published rules.

A run is scored whenever it ends with a termination code, including runs stopped by the elapsed-time cap; the engine settles the observing history accumulated so far. Of the 1\,224 runs selected for the published final ranking, 1\,115 reached the end of the campaign calendar, 41 reached the elapsed-time cap, 28 were ended by the submitted system and five ended with an execution error. The remaining 35 have no termination code in the public export, although they have scored results.

The entries ranked second and 73rd are marked as exceptions to the averaging rule. Their published task means are reproduced by selecting the three highest-scoring valid runs per card, from 15--21 and 24--31 candidates per card, respectively. All other ranked entries contain three scored runs per card. These selections reproduce all 408 task means and all 102 final scores.

Of the 1\,200 counted runs for the 100 regularly finalised teams, 427 have a non-zero replacement counter. In 103 of their 400 task groups the three counted runs have no common execution-time overlap; their median spread in start times is 9.3\,h. The metadata record replacement counts and execution times. Shared service conditions and changes between attempts can correlate or shift outcomes, as discussed in Section~\ref{sec:repeatability}.

The platform archives complete decision traces, exposure ledgers and logs in team-specific result repositories. Public campaign replays provide released action histories, catalogues and observing conditions. The score distributions and repeatability statistics reported here are reconstructed from the official per-run score export; the target-level diagnostics in Section~\ref{sec:localdiagnostic} use the archived local L3 campaign.

\subsection{Reference-campaign reconstruction}
\label{app:reference_reconstruction}
The L3 reference campaign uses the Python implementation of the baseline model with language-model advice disabled, default adaptive search settings and the card's 900-s normalised computation budget. The archived runner, policy and task inputs are identified by cryptographic checksums. The system receives the catalogue, configuration and sequential information through the standard interface and completes 37 nights with 957 exposures. Replaying the archived actions reproduces the messages, target responses, validity flags and score components. Reconstructing the retained target maxima also reproduces the completion fractions and right-ascension uniformity penalty.

For the programme comparison in Figure~\ref{fig:timeline}(d), we hold each archived exposure $e$ fixed and evaluate its additional weighted response under each programme $p$:
\begin{equation}
B_e(p)=\sum_{i\in\mathcal A_e}w_i g_{ie}\,[m_{ie}(p)-1],
\qquad p_e^\star\in\operatorname*{arg\,max}_{p}B_e(p),
\label{eq:programme_reference}
\end{equation}
where $\mathcal A_e$ contains the geometrically accepted assignments and $p$ ranges over DARK, BRIGHT and BACKUP. Raw responses, target weights and exposure geometry are unchanged. The prescribed matching bonuses are 20\%, 12\% and 6\%, respectively. The total weighted response of this fixed exposure is $B_e(p)+\sum_{i\in\mathcal A_e}w_i g_{ie}$. The second term is independent of programme, so maximising $B_e$ also maximises the exposure's weighted response. The frozen scorer reproduces the archived multipliers for all 11\,238 target--exposure records. Accounting for six-decimal response rounding gives 907 unique choices and 50 zero-response ties. The baseline model attains a maximum in 807 exposures, including the ties. This retrospective calculation uses the full simulated conditions and keeps the executed exposure sequence fixed.

The first earthquake begins at 03:45~UTC on December~3, initially multiplying the background efficiency by 0.164; the efficiency loss decays exponentially with a four-night time-scale. An instrumental fault starts at 07:00~UTC on December~7 with multiplier 0.443 and persists until a correct report at 07:00~UTC on December~15. The second earthquake begins at 06:15~UTC on December~16 with initial multiplier 0.649. These event and repair times define the stages in Table~\ref{tab:l3_phases}. Figure~\ref{fig:timeline}(b) shows nightly efficiency statistics, while panel~(c) marks the exact event times.

All three events act on instrument efficiency and preserve the programme band at fixed atmosphere and geometry. Atmospheric events can change the band independently: a site-wide haze event overlaps December~7--11. On December~9, the observed targets have $b_{ie}=0.494$--$0.576$, within BRIGHT. On November~23, $b_{ie}=0.614$--$0.847$, giving 525 DARK and 33 BRIGHT records. These counts follow the synthetic quality thresholds and provide checks on the configured band assignments.

Programme mismatches concentrate between the first earthquake and fault repair (Table~\ref{tab:l3_phases}). The baseline model's fallback and tie-breaking rules retain a dependence on the response scale (Appendix~\ref{app:reference_heuristics}). Its fault diagnostic requires recent evidence and applies an earthquake-related reporting hold and an additional decline criterion. The policy makes a false report on December~4, repairs the fault on December~15 and makes another false report on December~17. Both false reports fall within the free allowance, giving a total reporting reward of 100. Following repair, five of the remaining 165 programme choices fall below the fixed-exposure reference.

The data-loss notification at 04:02:30~UTC on December~3 invalidates 12 target records from one earlier exposure, reducing the completed-target count from 4681 to 4677. At termination, 7983 targets have surviving exposure records and 6474 are complete, including 434 of 435 required targets. Each observing request completes six of its eight targets within its window and earns 100. Exposure-action time includes 12.33~h in closed intervals and exceeds open time on 19 nights. Closed intervals consume simulated time and contribute zero response, explaining why the exposure bars in Figure~\ref{fig:timeline}(d) can exceed the open-time curve.

\begin{table*}[ht!]
\centering\small
\caption{Stages of the L3 campaign with the baseline model, separated by instrumental-event and repair times in Figure~\ref{fig:timeline}(c).}
\label{tab:l3_phases}
\begin{tabular}{lrrrrr}
\toprule
Stage & Exposures & Median $\eta$ & Median $T_e$ (s) & BACKUP choices & Below maximum bonus\\
\midrule
Before first earthquake & 582 & 0.946 & 600 & 4 & 18\\
Earthquake recovery & 90 & 0.473 & 300 & 39 & 53\\
Instrumental fault & 120 & 0.369 & 600 & 70 & 74\\
After fault repair & 10 & 0.930 & 2400 & 0 & 0\\
After second earthquake & 155 & 0.798 & 900 & 0 & 5\\
\bottomrule
\end{tabular}
\tablecomments{Exposures are grouped by start time; an exposure crossing a boundary remains in its starting stage. Efficiency medians use open environmental slots starting in each stage, including repair. The last two columns describe the baseline model. Below-maximum choices use equation~\ref{eq:programme_reference} with all other exposure properties fixed. The 50 zero-response ties (38 before the first earthquake and 12 during the fault) count as attaining the maximum.}
\end{table*}

\section{Published scores and representative systems}
\label{app:implementation_details}
\subsection{Detailed scores and illustrative systems}
\label{app:case_studies}

\begin{figure*}[ht!]
\centering

\includegraphics[width=0.985\textwidth]{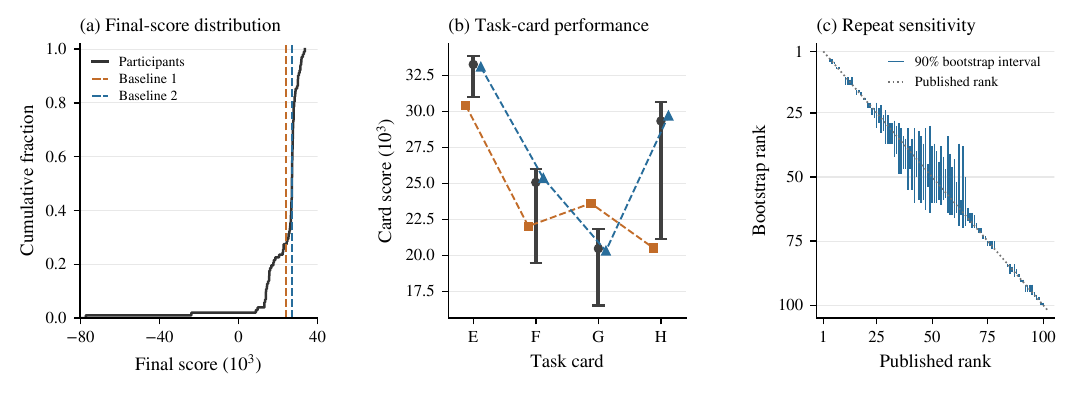}

\caption{Final results reconstructed from the official per-run score export. (a) Cumulative distribution of the 102 participant final scores, with the two baseline models. (b) Participant task medians and interquartile ranges (Weibull convention), compared with the same references. (c) 90\% rank intervals from 2000 resamplings of the three counted runs per team and task; the diagonal marks the published rank. For the two exceptional entries, resampling is conditional on their selected best three runs.}
\label{fig:results}\label{fig:scoring}

\alttext{Three panels show the final-score cumulative distribution, per-task medians and interquartile ranges with two baseline models, and bootstrap rank intervals against published rank.}
\end{figure*}

\begin{table*}[ht!]
\centering\footnotesize
\caption{Baseline models and the six highest-ranked submissions: per-task means and final scores reconstructed from the official run records.}
\label{tab:main}\label{tab:references}\label{tab:top10}
\begin{tabular}{lrrrrr}
\toprule
Entry & E & F & G & H & Final\\
\midrule
Baseline~1 & 30\,408 & 22\,015 & 23\,609 & 20\,502 & 24\,134\\
Baseline~2 & 33\,078 & 25\,346 & 20\,297 & 29\,683 & 27\,101\\
\midrule
1 Pleiades & 37\,543 & 31\,005 & 31\,412 & 34\,343 & 33\,576\\
2 starwing & 37\,194 & 30\,722 & 30\,307 & 33\,981 & 33\,051\\
3 MiXingJiao & 37\,312 & 30\,206 & 29\,441 & 33\,435 & 32\,599\\
4 CCB & 37\,056 & 29\,905 & 28\,207 & 32\,528 & 31\,924\\
5 Argelander & 36\,388 & 29\,716 & 27\,695 & 33\,432 & 31\,808\\
6 cloudbrain & 36\,969 & 29\,502 & 27\,216 & 33\,035 & 31\,680\\
\bottomrule
\end{tabular}
\tablecomments{Each task score is the mean of three runs, and the final score is their equal-weight mean over E--H. For starwing the counted runs are the three highest-scoring valid candidates per task (Appendix~\ref{app:execution_records}). Values are rounded only for display.}
\end{table*}

Table~\ref{tab:main} gives the task-level scores of the two baseline models and six leading teams.

\begin{figure*}[ht!]
\centering
\includegraphics[width=0.985\textwidth]{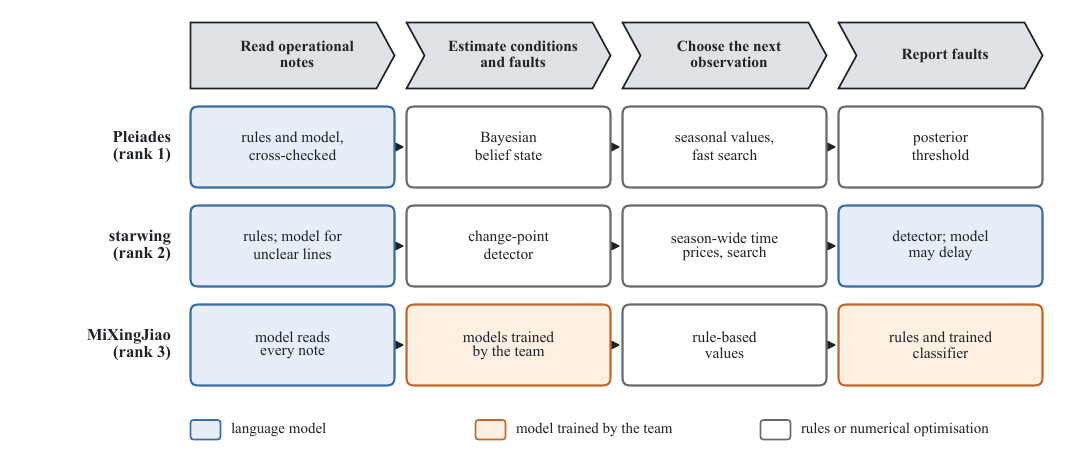}
\caption{{ The three highest-ranked decision systems, mapped onto four operations at each decision step. The colour of each box shows what does the work in that step: a language model, a model trained offline by the team, or rules and numerical optimisation. The descriptions follow the organisers' archived October 9 code-review report; rank labels refer to the published final leaderboard.}}
\label{fig:topagents}
\alttext{A grid with the three leading teams as rows and four steps as columns: reading operational notes, estimating conditions and faults, choosing the next observation and reporting faults; each cell names the method and is coloured by its type. }
\end{figure*}

Figure~\ref{fig:topagents} compares the three leading systems using the organisers' October 9 code-review report. The first maintains a belief state over observing conditions and faults. The second uses a seasonal linear programme to assign observing-time costs and a change-point detector for faults. The third combines a rule-based planner with two estimators trained offline and uses a language model to convert duty-log text into checked structured facts. These examples illustrate different ways to combine state estimation, numerical planning and operational information.

\section{Exploratory observing tasks}
\label{app:exploratory}
These tests use the frozen submissions of the 27 highest-ranked teams, with one run per tested configuration. Model-enabled runs route calls to a common model service. Paired runs block model access and invoke each system's existing fallback. The tests examine operational-text interpretation and adaptation to changes in survey conditions and priorities.

\subsection{Operational-text tests}
\label{app:text_tests}
\paragraph{Guide-camera failures (X).}
Task X adds noisy operational logs to the public text-heavy tasks. The logs announce guide-camera failures and normal calibration periods using three languages, varied time formats, corrections, rumours and unrelated messages. A third configuration adds the same information as a one-line structured summary. At fixed observing plan, reporting each genuine fault at its announced start time gives a score of approximately $+27\,700$; the same plan with fault reporting disabled gives approximately $-70\,400$. Parsers developed for A1--D1 score approximately $-68\,000$ to $-69\,000$ on the new wording.

For model-enabled submissions, the Spearman correlation with the final ranking is $\rho=0.82$. The original top ten have a median score of approximately $+30\,200$, compared with $+6\,700$ for ranks 11--27. Across 25 paired submissions, the median model-access difference $S_{\mathrm{on}}-S_{\mathrm{off}}$ is approximately $+3\,400$, and the median difference between the structured-summary and original model-enabled configurations is approximately $+200$. R7, R17 and R26 gain approximately 64\,000--72\,000 points with model access (Figure~\ref{fig:allteams}). The severe efficiency loss also produces a clear signature in exposure feedback: the six leading systems score 26\,000--33\,000 with model access blocked.

\paragraph{Subtle faults (Y) and revised notices (Z).}
Task Y announces a fibre-positioner service that lowers throughput by 15\% for 30\,min. Positioner self-tests produce the same throughput change and serve as decoys. Task Z announces atmospheric-dispersion-corrector services through engineering tickets. Tickets can be revised, postponed to the next open night, linked to earlier notices or withdrawn; informal operator messages add irrelevant information. Each task contains 96 announced service events and six background instrumental faults.

At fixed observing plan, correct interpretation of the service notices gains approximately 8700 points on Y and 8900 on Z relative to the control with fault reporting disabled. Reporting every throughput decrease loses more than 13\,000 points in these comparisons. Across the 27 submissions, the model-enabled rank correlations with the final evaluation are $\rho=0.63$ for Y and 0.53 for Z. Two systems make at least 60 correct fault reports on Y; all systems remain below this count on Z. The median model-access differences are approximately $+400$ and $-100$ points, respectively.

\subsection{Changes in climate, scientific priorities and scale}
\label{app:transfer_tests}
The transfer tasks provide their revised parameters in the public configuration. X1 adopts a monsoon climate. Y1 changes several scientific and scoring settings, including more than doubling the required-target fraction, multiplying its missed-target penalty by five and charging every false report. It also changes class weights, programme multipliers, uniformity and request settings. Z1 uses 40\,000 targets over approximately 9000~deg$^2$, with additional faults, data loss and pointing offsets.

For model-enabled runs, the rank correlations with the final evaluation are $\rho=0.85$, 0.38 and 0.87 for X1, Y1 and Z1, respectively. The largest rank changes occur under Y1's combined configuration, where large penalties for incomplete required targets accompany the largest declines.

\bibliographystyle{aasjournalv7.1}
\bibliography{references}

\end{document}